\documentclass[]{aa}
\usepackage[english]{babel}
\usepackage[utf8]{inputenc}
\usepackage[T1]{fontenc}
\usepackage{graphics,graphicx}
\usepackage[varg]{txfonts}
\usepackage{rotating}
\usepackage{lscape}
\usepackage{longtable}
\usepackage{amsmath,amssymb}
\usepackage{setspace}
\usepackage{booktabs}
\usepackage{graphicx,natbib}
\usepackage{epstopdf}
\usepackage{bm}
\usepackage{siunitx}
\usepackage{natbib}
\usepackage{eso-pic}
\usepackage{wallpaper}
\def\HII{H{\sc ii}}

\def\kms{\mbox{km~s$^{-1}$}}
\def\cmc{cm$^{-3}$}

\begin{document}
\title{Nitrogen and hydrogen fractionation in high-mass star forming cores from observations of HCN and HNC}
\author{L. Colzi\inst{1,2}
            \and
            F. Fontani\inst{2} 
            \and
            P. Caselli\inst{3}
            \and 
            C. Ceccarelli\inst{4,5}
            \and
            P. Hily-Blant\inst{4,5}
            \and 
            L. Bizzocchi\inst{3} 
            }
\institute{  Università  degli studi di Firenze, Dipartimento di fisica e Astronomia, Via Sansone, 1 -50019 Sesto Fiorentino (Italy)
           \and 
           INAF-Osservatorio Astrofisico di Arcetri, Largo E. Fermi 5, I-50125, Florence, Italy
           \and
           Max-Planck-Instit\"{u}t f\"{u}r extraterrestrische Physik, Giessenbachstrasse 1, D-85748, Garching bei M\"{u}nchen, Germany
           \and
           CNRS, IPAG, F-38000 Grenoble (France)
           \and
          Univ. Grenoble Alpes, IPAG, F-38000 Grenoble (France)
           }
\date{Received date; accepted date}

\titlerunning{Nitrogen and hydrogen fractionation}
\authorrunning{Colzi et al.}

 
\abstract{The ratio between the two stable isotopes of nitrogen, $^{14}$N and $^{15}$N,
is well measured in the terrestrial atmosphere ($\sim 272$), and for the pre-Solar nebula
($\sim 441$, deduced from the Solar wind). 
Interestingly, some pristine Solar System materials show enrichments in $^{15}$N 
with respect to the pre-Solar nebula value. However, it is not yet clear if, and how, these 
enrichments are linked to the past chemical history, due to the limited number of measurements 
in dense star-forming regions. In this respect, dense cores believed to be precursors of
clusters containing also intermediate- and high-mass stars are important targets, 
as the Solar System was probably born within a rich stellar cluster.
The number of observations in such high-mass dense cores has remained limited so far.
In this work, we show the results of IRAM-30m observations of the 
\emph{J}=1-0 rotational transition of the molecules HCN and HNC, and their $^{15}$N-bearing 
counterparts, towards 27 intermediate/high-mass dense cores divided almost equally
in three evolutionary categories: high-mass starless cores, high-mass protostellar objects,
and ultra-compact \HII\ regions. We have also observed the DNC(2-1) rotational
transition, in order to search for a relation between the isotopic ratios
D/H and $^{14}$N/$^{15}$N. We derive average $^{14}$N/$^{15}$N ratios of $359\pm16$ 
in HCN and of  $438\pm21$ in HNC, with a dispersion of about 150-200. We find no trend of the
 $^{14}$N/$^{15}$N ratio with the evolutionary stage. This result agrees with what found
from N$_{2}$H$^{+}$ and its isotopologues in the same sources, although the $^{14}$N/$^{15}$N 
ratios from N$_{2}$H$^{+}$ show a dispersion larger than that in HCN/HNC.
Moreover, we have found
no correlation between D/H and $^{14}$N/$^{15}$N in HNC. These findings indicate 
that: (1) the chemical evolution does not seem to play a role in the fractionation of
nitrogen; (2) the fractionation of hydrogen and nitrogen in these objects are not related.
}
  

\maketitle

 \section{Introduction}
 \label{intro}
 Nitrogen is the fifth most abundant element in the Universe. It possesses
 two stable isotopes, $^{14}$N and $^{15}$N. F\"uri \& Marty (\citeyear{furi}) proposed that there are three distinct isotopic reservoirs in the Solar system: the
protosolar nebula, the inner Solar system and cometary ices. In the terrestrial atmosphere (TA), 
 the typical isotopic ratio $^{14}$N/$^{15}$N, as derived from N$_{2}$, is $\sim$272 (Marty et al.~\citeyear{marty09}). 
 This value is almost a factor two larger than the ratio measured in nitrile-bearing 
 molecules and in nitrogen hydrides of some comets (e.g.~$\sim$150, Manfroid et al.~\citeyear{manfroid}, Shinnaka et al.~\citeyear{shinnaka}), and larger of even a factor five in carbonaceus chondrites (e.g.~$\gtrsim$50, Bonal et al.~\citeyear{bonal}).
 On the other hand, the ratio measured in Solar wind particles collected by the Genesis spacecraft, representative of the proto-Solar nebula value (PSN) and similar to that measured on Jupiter (Owen et al.~\citeyear{owen}), is 441$\pm$6 (Marty et al.~\citeyear{marty}). Thus, the PSN $^{14}$N/$^{15}$N value is about 2 times larger than the TA value and 3 times larger than the value measured in comets. These measurements suggest that multiple isotopic reservoirs were present very early in the formation process of the solar system. This was recently demonstrated based on the CN/C$^{15}$N and HCN/HC$^{15}$N isotopic ratios measured in a sample of PSN analogs (Hily-Blant et al.~\citeyear{hily-blant17}). The nature and origin of these reservoirs remain, however, elusive. Currently, the two main possibilities are i) isotope-selective photodissociation of N$_{2}$ in the PSN by stellar UV (Heays et al.~\citeyear{heays}, Guzman et al.~\citeyear{guzman}), and ii) an interstellar origin by mass fractionation reaction (Hily-Blant et al.~\citeyear{hily-blant13b}, Roueff et al.~\citeyear{roueff}).
  A similar isotopic enrichment in comets, some meteorites, and Interplanetary Dust Particles 
 with respect to the PSN value has been found also for hydrogen. In the PSN, the D/H ratio is $\sim$10$^{-5}$ (Geiss et al.~\citeyear{geiss}),
 i.e. similar to the cosmic elemental abundance (Linsky et al.~\citeyear{linsky}).
 In comets, different values of the D/H ratio were estimated: in the Jupiter family 
 comet 67P/Churyumov-Gerasimenko the D/H ratio measured in H$_{2}$O is about 5$\times$10$^{-4}$,
 approximately three times that of Earth's oceans (Altwegg et al.~\citeyear{altwegg}), and other results, from Herschel, have shown D/H$\sim$ 1.5$\times$10$^{-4}$ 
 in another Jupiter family comet 103P/Hartley (Hartogh et al.~\citeyear{hartogh}) same as the Earth's ocean.
 In carbonaceous chondrites, values of D/H$\sim$1.2-2.2$\times$10$^{-4}$ in hydrous silicates were 
 obtained (Robert \citeyear{robert}). Furthermore, very high D/H ratios of 
 $\sim$10$^{-2}$ (Remusat et al.~\citeyear{remusat}) have been found in small 
 regions in Insoluble Organic Matter (IOM) of meteorites and Interplanetary Dust 
 Particles (IDPs), called "hot spots". Deuterium fractionation in the Interstellar Medium and in the Solar System objects was also discussed by Ceccarelli et al.~(\citeyear{ceccarelli14}). 

 From a theoretical point of view, for several species the D/H enhancement has 
 its origin in a low-temperature environment where ion-molecule reactions are favoured, starting from the reaction:
 \[
 {\rm H}_{3}^{+}+{\rm HD} \rightarrow {\rm H}_{2}{\rm D}^{+}+{\rm H}_{2}+{\rm 232K} ,
 \]
which produces an enhanced H$_2$D$^+$/H$_{3}^{+}$ abundance ratio, for temperatures lower than $\sim$50 K and when H$_{2}$ is mainly in para form (e.g. Pagani et al.~\citeyear{pagani}; Gerlich et al~\citeyear{gerlich}; Walmsley et al.~\citeyear{walmsley}). Moreover,
in dense cores, where CO freezes-out on dust grains (e.g.~Caselli et 
al.~\citeyear{caselli99}, Fontani et al.~\citeyear{fontani12}), H$_{2}$D$^{+}$ can 
survive and H$_2$D$^+$/H$_{3}^{+}$ is further enhanced causing a high level of deuteration in molecules. The role of 
grain surface chemistry on icy material during the early cold phase is also expected
to play an important role for 
the deuteration of neutral species like water, formaldehyde, methanol and complex organic molecules
(e.g. Cazaux et al.~\citeyear{cazaux}, ~Taquet et al.~\citeyear{taquet12},~\citeyear{taquet13}). This is due to the fact the enhanced abundances of the deuterated forms of H$_{3}^{+}$ (H$_{2}$D$^{+}$, D$_{2}$H$^{+}$, D$_{3}^{+}$) produce enhanced abundances of D atoms in the gas phase, upon their dissociative recombination with electrons. The enhanced D abundance implies a larger D/H ratio, so that unsaturated molecules on the surface (in particular CO) can be deuterated as well as hydrogenated to produce singly and doubly deuterated water and formaldehyde as well as singly, doubly and triply deuterated methanol (e.g. Caselli \& Ceccarelli \citeyear{caselliceccarelli}; Ceccarelli et al.~\citeyear{ceccarelli14}).

In principle, similar gas-phase mass fractionation reactions can also produce $^{15}$N enrichments under cold and dense conditions. This motivated the search for correlated enrichments in D and $^{15}$N in cosmomaterials (Aléon et al.~\citeyear{aleon}) although chemical models suggest that such correlations may not be present (Wirstr\"{o}m et al.~\citeyear{wirstrom}). 
Furthermore, while the above scenario of D-enrichments through ion-neutral mass fractionation reactions has been proved firmly for D through 
observations of low- and high-mass star-forming regions (e.g.~Crapsi et 
al.~\citeyear{crapsi}, Caselli et al.~\citeyear{caselli08}, Emprechtinger et al.~\citeyear{emprechtinger}, Fontani et 
al.~\citeyear{fontani11}), the reasons for $^{15}$N enrichment are still
 highly uncertain. For example, HC$^{15}$N and H$^{15}$NC could be formed through the dissociative recombination of HC$^{15}$NH$^{+}$: in fact, Terzieva \& Herbst (\citeyear{terzieva}) found that the reaction that causes the most of N-fractionation is the exchange reaction between $^{15}$N and HCNH$^{+}$:
\[
 ^{15}{\rm N} + {\rm HCNH}^{+} \rightarrow {\rm N} + {\rm HC}^{15}{\rm NH}^{+} + 35.1{\rm K},
 \]
but they assumed that this reaction could occur without an energy barrier. The most recent and complete
 chemical models (Roueff et al.~\citeyear{roueff}) 
 are implementend with the recent discovery that this reaction have an energy barrier and indicate that $^{15}$N should not be enriched during the evolution of a
 star-forming core, not even at the very early cold phases. Moreover, the observational 
works devoted to test these predictions are still very limited.
 
 In the few low mass pre-stellar cores or protostellar envelopes observed so far, 
values of $^{14}$N/$^{15}$N comparable to the value of the PSN have been 
measured from N$_{2}$H$^{+}$ and NH$_{3}$ (330$\pm$150 in N$_{2}$H$^{+}$, Daniel et al.~\citeyear{daniel}; 350 - 500, Gerin et 
al.~\citeyear{gerin}; 334$\pm$50, Lis et al.~\citeyear{lis} in NH$_{3}$), or even larger (1000$\pm$200 in N$_{2}$H$^{+}$, Bizzocchi et 
al.~\citeyear{bizzocchi}). Conversely, through observations of the 
 nitrile-bearing species HCN e HNC in low-mass sources, the $^{14}$N/$^{15}$N ratio turns out to be 
 significantly lower (140 - 360, Hily-Blant et al.~\citeyear{hily-blant13a}; 160-290, Wampfler et al.~\citeyear{wampfler}). Moreover low values have been recently found also in protoplanetary disks (from 80 up to 160, Guzm\'{a}n et al.~\citeyear{guzman}), but this results are based on a low statistics (six protoplanetary disks).
From the point of view of chemical models, differential fractionation is expected between nitriles and hydrides (Wirstr\"{o}m et al.~\citeyear{wirstrom}, Hily-Blant et al.~\citeyear{hily-blant13a},~\citeyear{hily-blant13b}), although the most recent models (Roueff et al.~\citeyear{roueff}) do not support this scenario. However, none of these models is able to reproduce the low fractionation observed in N$_{2}$H$^{+}$ towards L1544 (Bizzocchi et al.~\citeyear{bizzocchi}) and the large spread in Fontani et al.~\citeyear{fontani15a}.
 
To investigate the possible correlation between D and $^{15}$N enrichments in interstellar environments,  Fontani et al.~(\citeyear{fontani15a}) conducted a survey of the $^{14}$N/$^{15}$N and D/H isotopic ratios towards a sample of high-mass cores in different evolutionary stages and found a possible anticorrelation, which, however,
was quite faint and obtained from a low number (12 objects) of detections.
In addition, no correlation between the disk averaged D/H and $^{14}$N/$^{15}$N 
ratios have been measured by Guzm\'{a}n et al.~(\citeyear{guzman}) in six protoplanetary disks, but again the conclusion is based on a poor statistics. 
Another intriguing result is that the D- and $^{15}$N-enhancements 
 are not always observed in the same place in pristine Solar System material (Busemann et al.~\citeyear{busemann}; Robert \& Derenne \citeyear{robert06}).
Therefore, it is important to gather more data in sources that are good
candidates to represent the environment in which our Sun was born to put stringent 
constraints on current chemical models.
In this respect, intermediate- and high-mass star-forming cores are interesting targets
because growing evidence is showing that the Sun was born in a rich cluster, containing 
also massive stars (Adams \citeyear{adams}, Banerjee et al.~\citeyear{banerjee}). Moreover, Taquet et al.~(\citeyear{taquet16})
have recently proposed that the proto-Sun was born in an environment denser and warmer
than that usually considered as a Solar System progenitor. In any case, because the statistic is still poor, having more
observations in star-forming cloud cores in different evolutionary stages is useful to 
better understand if and how the $^{15}$N fractionation process is eventually influenced by
evolution. An example of existing work about $^{15}$N fractionation in high-mass dense cores is this of Adande \& Ziurys (\citeyear{adande}). This work have a larger beam with respect to that used in our observations and they do not have an evolutionary classification of the sources, so their results are difficult to be interpreted in an evolutionary study.
 
In this work we report the first measurements of the $^{14}$N/$^{15}$N ratio 
derived from HCN and HNC in a sample of 27 dense cores associated with different stages 
of the high-mass star formation process, and already studied in deuterated molecules 
and in the $^{15}$N-bearing species of N$_{2}$H$^{+}$ (Fontani et al.~\citeyear{fontani11}, 
\citeyear{fontani15a}, \citeyear{fontani15b}). 
In particular, Fontani et al.~(\citeyear{fontani15a}) have measured, for the first time, the 
$^{14}$N/$^{15}$N isotopic ratio in N$_{2}$H$^{+}$ towards the same sources.
Therefore, these new data allow us to investigate the possible difference between  nitrogen hydrides and nitrile-bearing species proposed by both theoretical studies (Wirstr\"{o}m 
et al.~\citeyear{wirstrom}) and observational findings (Hily-Blant et al.~\citeyear{hily-blant13a}).
We report also measurements of the D/H ratios for HNC, to search for relations between the two isotopic ratios.

\section{Observations}
\label{obs}

We performed observations of the \emph{J}=1-0 rotational transition of H$^{15}$NC, 
HN$^{13}$C, HC$^{15}$N and H$^{13}$CN towards the 27 sources observed by
Fontani et al.~(\citeyear{fontani15a}) from 6 to 9 June, 2015, using the 3~mm receiver
of the IRAM-30m telescope. We refer to this paper for the description of the 
source sample. We simultaneously observed the \emph{J}=2-1 transition of DNC 
with the 2 mm receiver. Table \ref{observation} presents the observed spectral windows and
some main technical observational parameters. Table \ref{h13cn-iperf} presents the hyperfine frequencies of H$^{13}$CN. The atmospheric conditions 
were very stable during the whole observing period, with precipitable water 
vapour usually in the range 3 -- 8~mm. 
The observations were made in wobbler-switching mode with a wobbler throw of 240''. Pointing was 
checked almost every hour on nearby quasars, planets, or bright \HII\ regions. 
The data were calibrated with the chopper wheel technique (see Kutner \& Ulich \citeyear{kutner}), 
with a calibration uncertainty of about 10\% . 
The spectra were obtained in antenna temperature units, T$_{\rm A}^{*}$, and then converted to 
main beam brightness temperature, T$_{\rm MB}$, via the relation $T_{\rm A}^{*}=T_{\rm MB}(B_{eff}/F_{eff})=T_{MB} \eta_{\rm MB}$, where $\eta_{\rm MB}=B_{eff}/F_{eff}$ is the ratio between the Main Beam efficiency ($B_{eff}$) of the telescope and the Forward efficiency of the telescope ($F_{eff}$). The spectra were 
obtained with the fast Fourier transform spectrometers with the
finest spectral resolution (FTS50), providing a channel width of 50~kHz. 
All calibrated spectra were analysed using the GILDAS\footnote{The GILDAS
software is available at http://www.iram.fr/IRAMFR/GILDAS} software developed at 
the IRAM and the Observatoire de Grenoble. Baselines in the spectra were all fitted 
by constant functions, or polynomials of order 1. The rest frequencies used for 
the line identification, have been taken from different laboratory works: HC$^{15}$N from Cazzoli et al.~(\citeyear{cazzoli}), H$^{13}$CN from Cazzoli \& Puzzarini (\citeyear{cazzolipuzzarini}), HN$^{13}$C from van der Tak et al.~(\citeyear{vandertak}), H$^{15}$NC from Pearson et al.~(\citeyear{pearson}) and DNC from Bechtel et al.~(\citeyear{bechtel}). The other spectroscopic parameters used in the derivation of the column densities have been taken from the Cologne Molecular Database for Spectroscopy\footnote{http://www.astro.uni-koeln.de/cdms} (CDMS; M\"{u}ller et al.~\citeyear{muller01},~\citeyear{muller05}) except for H$^{15}$NC, for which we have used the Jet Propulsion Laboratory database\footnote{https://spec.jpl.nasa.gov/}.

\begin{table}
\tiny
\caption{Line rest frequencies and observational parameters}
  \begin{tabular}{lllllll}
  \hline
  Line   & Frequency & HPBW & $\Delta v^{*}$& $T_{\rm sys}$  & $\eta_{\rm MB}$ &$F_{eff}$\\ 
           &    GHz               &        arcsec           &        \kms                &       K             &     & \\
  \hline
HC$^{15}$N(1-0)  & 86.0549  &   28      &  $\sim 0.16$ & 100--150 & 0.85 &0.95 \\
H$^{13}$CN(1-0)  & 86.3399  &   28      &  $\sim 0.16$ & 100--150 & 0.85  &0.95\\
HN$^{13}$C(1-0)  & 87.0908  &   28      &  $\sim 0.16$ & 100--150 & 0.85  &0.95\\
H$^{15}$NC(1-0)  & 88.8657 &   27      &  $\sim 0.16$ & 100--150 & 0.84  &0.95\\
DNC(2-1)              & 152.60977  & 15     &  $\sim 0.096$ & 200--250  &  0.78 & 0.93\\
  \hline
  \normalsize
  \label{observation}
  \end{tabular}

  *: velocity resolution of the spectrum.
\end{table}

\begin{table}[t]
\tiny
 \caption{Frequencies of the hyperfine components of the transition H$^{13}$CN(1-0). F is the quantum number associated with the sum between the orbital angular momentum |J| and the $^{14}$N nuclear angular momentum. The hyperfine spitting due to $^{13}$C is negligible.}
 \begin{tabular}{ccc}
  \toprule
  & &  Frequency\\
  J'-J'' & F'-F'' & (GHz)\\
  \midrule
  1-0 &1-1 &$86.33873$\\
   &2-1  &$86.34016$\\
  &0-1  &$86.34225$\\
\bottomrule
 \end{tabular}
 \centering
\label{h13cn-iperf}
\normalsize
\end{table}
 \section{Results}
 \label{res}
 \subsection{H$^{15}$NC, HN$^{13}$C, HC$^{15}$N and H$^{13}$CN}
 
The H$^{15}$NC(1-0) line has been detected in 26 cores (96.3$\%$): 11 HMSCs, 8 HMPOs and 
7 UC HIIs; the HC$^{15}$N(1-0) line has been detected in 24 cores (88.8$\%$): 8 HMSCs, 
9 HMPOs and 7 UC HIIs; the HN$^{13}$C(1-0) line is detected in all sources (100$\%$)
and also the H$^{13}$CN(1-0) line is clearly detected in all sources (100$\%$).

First of all, to evaluate the isotopic ratios we have used the $^{13}$C-bearing
species of HCN and HNC because the main isotopologues are usually optically 
thick (e.g.~Padovani et al.~\citeyear{padovani}). Then, we can derive the 
$^{14}$N/$^{15}$N ratio by 
correcting for the $^{12}$C/$^{13}$C ratio. This latter has been derived from 
the relation between this ratio and the source galactocentric distance found 
for CN by Milam et al.~(\citeyear{milam05}). Galactocentric distances have 
been taken from Fontani et al.~(\citeyear{fontani14}, \citeyear{fontani15a}).

Both H$^{15}$NC(1-0) and HC$^{15}$N(1-0) do not have hyperfine structure, 
then all the lines were fitted with a single Gaussian. On the contrary, HN$^{13}$C(1-0) 
and H$^{13}$CN(1-0) have hyperfine structure. This cannot be resolved for the 
HN$^{13}$C(1-0) because the line widths are always comparable to (or larger than) 
the separation in velocity of the hyperfine components. Then all the lines were fitted 
with single Gaussians too, and since the column densities will be derived
from the total integrated area of the rotational line, this simplified approach will not 
affect our measurements as long as the lines are optically thin, which is discussed in this Section. On the other hand, it can overestimate the intrinsic 
line width. To estimate of how much, we have fitted a line both with the Gaussian 
and the hyperfine method, and we have found that with a single Gaussian, the line 
width is about 30$\%$ larger than that obtained with the hyperfine structure. 

Moreover, the fit using the hyperfine method have demonstrated that the lines are optically thin. Finally, for H$^{13}$CN(1-0), for which we could resolve
the hyperfine structure, we fitted the components simultaneously assuming
that they have the same excitation temperature (T$_{\rm ex}$) and line width, and that the
separation in velocity is fixed to the laboratory value. 
This fitting procedure gave us good results in all spectra. 
The Gaussian fitting results are listed in Tab.~\ref{hn13c-h15nc-fit} and \ref{hc15n-dnc-fit} 
and the hyperfine fitting results are listed in Tab.~\ref{h13cn-fit}. 
In Fig.~\ref{hn13c}, \ref{h13cn}, \ref{h15nc} and \ref{hc15n} we show the spectra
of the H$^{15}$NC, HC$^{15}$N, HN$^{13}$C and H$^{13}$CN(1-0) lines for all the 27 sources. As can be noted, the hyperfine components of H$^{13}$CN(1-0) were always detected. The spectra of AFGL5142-EC may be partially contaminated from the nearby core AFGL5142-mm, but this emission is expected to be not dominant because the angular separation between the two cores is 30'' (Busquet et al.~\citeyear{busquet}), equal than the beam of the telescope.

The total column densities of the four species, averaged within the beam, have been 
evaluated from the total line integrated intensity using eq.~(A4) of Caselli et al.~(\citeyear{caselli02}),
which assumes that T$_{\rm ex}$ is the same for all transitions within the same molecule, and 
optically thin conditions. The assumption of optically thin lines
is justified by the fact that from all the hyperfine fits of H$^{13}$CN(1-0) we find $\tau\ll$1 
and well constrained ($\Delta\tau$/$\tau\leq$1/3). We assume that also the lines of the other 
isotopologues are optically thin. We assume LTE conditions, as all the observed
sources have average H$_2$ volume densities of the order of $\sim 10^5$ \cmc\ (see
Fontani et al.~\citeyear{fontani11} and references therein), i.e.
comparable or marginally smaller than the critical densities of the lines observed, thus
this assumption is also reasonable. Because the T$_{\rm ex}$ cannot be deduced from our optically thin spectra (and cannot deduced T$_{\rm ex}$ also because we have only one transition), we adopted as T$_{\rm ex}$ the kinetic temperatures given by Fontani et al.~(\citeyear{fontani11}), who derived them following the method described in Tafalla et al.~(\citeyear{tafalla}) based on Monte Carlo models from which is obtained a relation among the kinetic temperature, T$_{k}$, and the NH$_{3}$ rotation temperature between metastable levels. The T$_{k}$ values are given 
in the last column of Table~\ref{risultati1} and \ref{risultati2}. This last assumption is
critical for the single column densities, but the HN$^{13}$C/H$^{15}$NC and 
H$^{13}$CN/HC$^{15}$N column density ratios do not change significantly by 
varying T$_{k}$ between 20 and 100~K: changes are of one per cent or of ten 
per cent depending on the source. All the column densities and the parameters 
used to derive them (line integrated intensities) are given in Tables~\ref{risultati1} 
and Tab.~\ref{risultati2}. We consider as detections the lines with
$T_{MB}^{peak}\geq 3\sigma$.
For the lines not clearly detected,  we have distinguished between those with a 
$2.5\sigma \leq T_{MB}^{peak}< 3\sigma$, and those with 
$T_{MB}^{peak}<2.5\sigma$. For the first ones, considered as tentative detections, 
we have computed the total column densities as explained above, and for the latter 
ones we have given an upper limit to the integrated areas, and hence to the total 
column densities using:
\[
\int T_{MB}\,dv=\frac{\Delta v_{1/2}T_{MB}^{peak}}{2\sqrt{\frac{ln2}{\pi}}},
 \]
 where $\sigma$ is the r.m.s. of the spectra, $T_{MB}^{peak}$ is taken equal to 3$\sigma$ and $\Delta v_{1/2}$ is the average value of the FWHM of the lines clearly detected for the corresponding transition and evolutionary stage of the source. The average value $\Delta v_{1/2}$ for the high-mass starless cores in our data is $\Delta v_{HMSC}=2.2\pm 0.3$ km/s, while for the high-mass protostellar objects is $\Delta v_{HMPO}=1.8\pm 0.2$ km/s.
Finally we have derived the column density uncertainties from the errors on the line areas for optically thin lines, given by $\sigma\times\Delta v\times\sqrt{N}$ ($\sigma$= root mean square noise in the spectrum, $\Delta v$= spectral velocity resolution, $N$= number of channels with signal) and taking into account the calibration error ($10\%$) for the T$_{MB}$. Conversely, uncertainties in the $^{14}$N/$^{15}$N ratios have been computed from the propagation of errors on the column densities, as explained above, without taking the calibration uncertainties into account because the lines were observed in the same spectrum
(see Sect.~\ref{obs}), so that the calibration error cancels out in their ratio.

\subsection{DNC}
We have also detected the rotational transition DNC(2-1) for all the 27 sources.
These lines will be used to measure the D/H ratio, that will be compared with the 
$^{14}$N/$^{15}$N ratio. 
Such a high detection rate indicates that deuterated gas is present at every stage 
of the massive star and star cluster formation process, as it has already
been noted by Fontani et al.~(\citeyear{fontani11}). The transition possesses a
hyperfine structure that, because of the broad line widths 
(see Table \ref{hc15n-dnc-fit}), is not resolved. Therefore, the lines where 
fitted with a single Gaussian. To estimates how much the line widths are overestimated using Gaussian fits, we have fitted a line both with the Gaussian and the hyperfine method, and we have found that with a single Gaussian, the line width is about 10$\%$ larger than that obtained with the hyperfine structure.
We point out that for evolved sources (HMPOs and UC HIIs) there is another 
line in the spectra partly overlapping with DNC(2-1), identified as acetaldehyde 
at 152.608 GHz (the \emph{J}$_{Ka,Kc}$ = 8$_{0,8}$-7$_{0,7}$ transition). The fact that the line is detected
only in the evolved objects is consistent with the idea that acetaldehyde 
is probably released from grain mantles because it is detected only 
in the warmer and more turbulent objects (see e.g.~Codella et
al.~\citeyear{codella}). When we have fitted the DNC lines with Gaussians, 
we have excluded the contribution of this line by fitting the two lines
simultaneously and when possible we have just excluded this line from the fit. Fitting results are in Tab.~\ref{hc15n-dnc-fit}.

Also in this case we have determined the column densities under the assumption 
of optically thin conditions and same T$_{\rm ex}$ for all transitions using 
eq.~(A4) of Caselli et al.~(\citeyear{caselli02}):
\begin{equation} 
N_{TOT}=\frac{8\pi\nu_{ij}^{3}}{c^{3}A_{ij}}\frac{1}{g_{i}}\frac{W}{(J_{\nu_{ij}}(T_{ex})-J_{\nu_{ij}}(T_{BG}))}\frac{1}{\biggl(1-\exp{(-\frac{h\nu_{ij}}{kT_{ex}})}\biggr)}\frac{Q(T_{ex})}{\exp{(-\frac{E_{j}}{kT_{ex}})}}
\end{equation}
 where $\nu_{ij}$ is the frequency of the transition, $A_{ij}$ is the Einstein coefficient of  spontaneous emission, $g_{i}$ is the statistical weight of the upper level, $E_{j}$ is the energy of the lower level, $c$ is the speed of light, $k$ is the Boltzmann constant, $T_{\rm ex}$ is the excitation temperature of the transition,
$Q(T_{\rm ex})$ is the partition function at temperature $T_{\rm ex}$, and $T_{BG}$ is the background temperature (2.7 K), $J_{\nu_{ij}}(T)$ is
 \[
 J_{\nu_{ij}}(T)=\frac{h\nu_{ij}}{k}\frac{1}{\exp{(\frac{h\nu_{ij}}{kT})}-1},
 \]
W is the integrated intensity of the line ($\int T_{B}\,dv=(\int T_{MB}\,dv)\frac{\Omega_{MB}}{\Omega_{s}}$, where T$_{B}$ is the brightness temperature, $\Omega_{MB}$ is the solid angle of the main beam, and $\Omega_{s}$ is the solid angle of the source).
However, in this case the HNC/DNC ratio depends on the temperature because 
of the different excitation conditions of the two transitions observed ((2--1) for DNC 
and (1--0) for HN$^{13}$C), so that the ratio depends on the temperature by 
the factor $\exp({E_{j}/kT})$. 
We had also to correct the HNC/DNC ratios for the different beam of the antenna at the frequencies of the two lines and we have assumed that the emissions of DNC(2-1) and HNC(1-0) are less extended than the beam size of DNC(2-1). This last correction results in a factor 3.09 to be multiplied to the HNC/DNC ratio: \[
 \biggl(\frac{1.22\frac{\lambda_{1}}{D}}{1.22\frac{\lambda_{2}}{D}}\biggr)^{2}=\biggl(\frac{\lambda_{1}}{\lambda_{2}}\biggr)^{2}=\biggl(\frac{\nu_{2}}{\nu_{1}}\biggr)^{2}=3.09
\]
where $\lambda_{1}$ and $\lambda_{2}$ are the wavelenghts of the HNC(1-0) and DNC(2-1) transitions, respectively (and $\nu_{1}$ and $\nu_{2}$ the corresponding frequencies).
In Fig.~\ref{dnc}, the spectra of DNC(2-1) for all the 27 sources are shown. 
The total column densities are listed in Tab.~\ref{risultati1}.
Finally we have derived the errors as explained in Sect.~3.1, but here for the D/H ratios we have considered also the calibration uncertainties because the two lines were observed in separate setups.

\section{Discussion}
\label{disres}
\subsection{$^{15}$N-Fractionation}
\begin{figure}[htpb!]
\includegraphics[width=40pc]{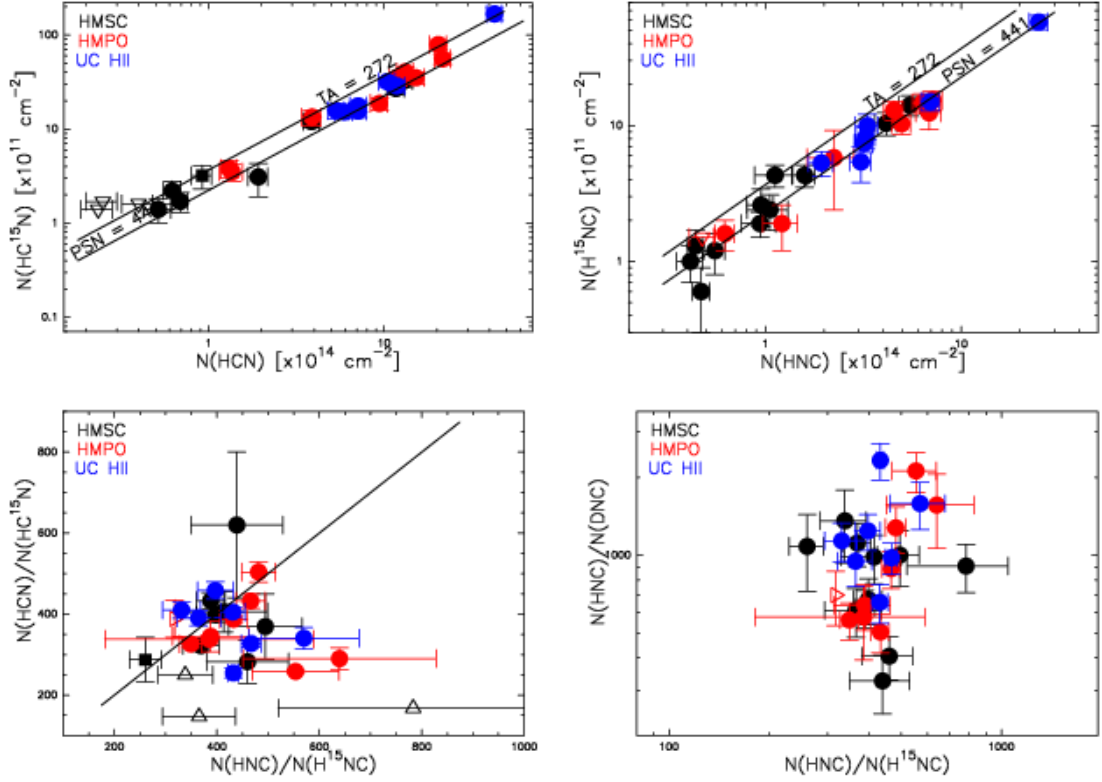}
\caption{Top panels: column density of HCN against that of HC$^{15}$N (left) and of HNC against that of H$^{15}$NC (right). Bottom
panels: comparison between the $^{14}$N/$^{15}$N isotopic ratios derived from the column density ratios N(HCN)/N(HC$^{15}$N) and N(HNC)/N(H$^{15}$NC) (right) and comparison
between the H/D and $^{14}$N/$^{15}$N isotopic ratios in HNC (left).
In all panels, the filled circles represent the detected sources (black=HMSCs; red=HMPOs; blue=UC HIIs). The open
triangles in the top panels are the upper limits on N(H$^{15}$NC) (left) and N(HC$^{15}$N) (right), while in the bottom panels the open triangles indicate
lower limits on either N(HCN)/N(HC$^{15}$N) or N(HNC)/N(H$^{15}$NC). The filled squares represent tentative detections. The solid lines
in the top panels indicate the mean atomic composition as measured in the terrestrial atmosphere (TA) and in the protosolar nebula (PSN), while in the
bottom-left panel the solid line indicate the locus of points where N(HCN)/N(HC$^{15}$N) is equal to N(HNC)/N(H$^{15}$NC).}
\centering
\label{15-N}
\end{figure}

The comparison between the column densities of the $^{15}$N-containing 
species and those of their main isotopologues, derived as explained in 
Sect.~\ref{res}, is shown in Fig.~\ref{15-N}. The corresponding 
$^{14}$N/$^{15}$N ratios are given in Tab.~\ref{risultati1} and in Tab.~\ref{risultati2}.

\begin{figure}[htpb!]
\centering
\includegraphics[width=45pc]{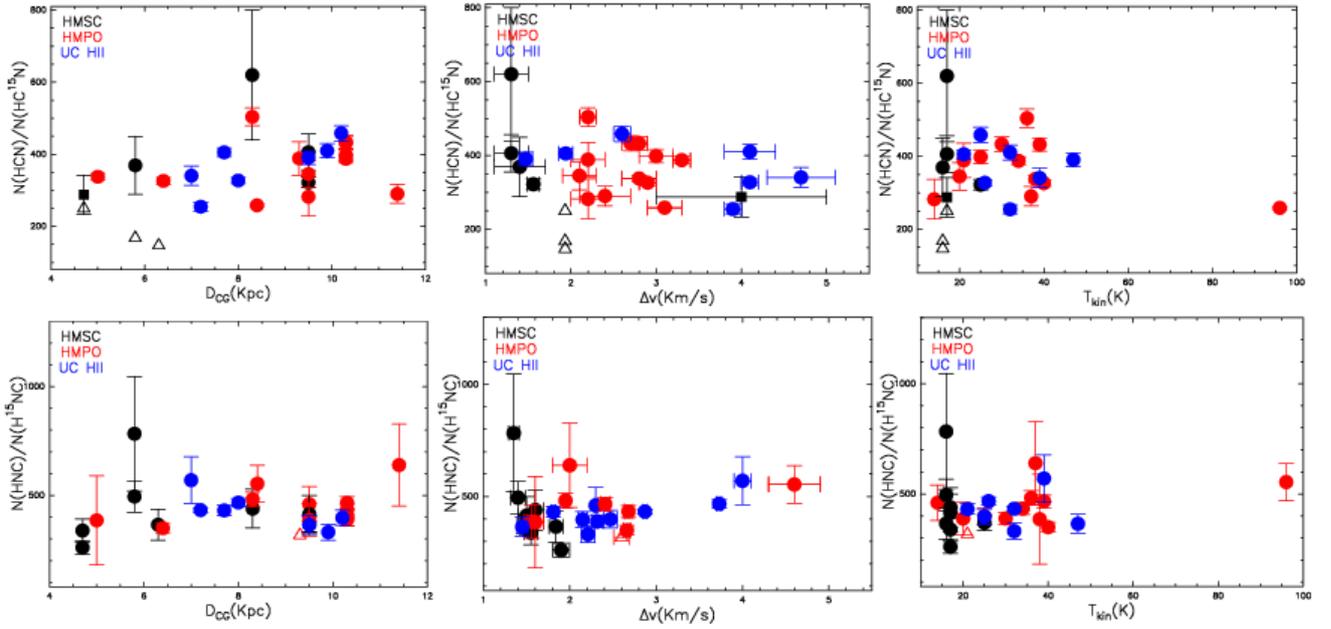}
\caption{Top panels: N(HCN)/N(HC$^{15}$N) as a function of Galactocentric distances, line widhts and kinetic temperatures.
Bottom panels: N(HNC)/N(H$^{15}$NC) as a function of Galactocentric distances, line widhts and kinetic temperatures.
The symbolism is the same used for bottom panels of Fig.\ref{15-N}.}
\centering
\label{15-NAndamenti}
\end{figure}

Let us first discuss the $^{15}$N-fractionation found for HCN. 
As we can see from the top left panel in Fig.~\ref{15-N}, there is not a large
spread of measured values, which are very close to the value found for the PSN. 
The mean values for the three evolutionary stages are: 346$\pm$37 for HMSCs, 
363$\pm$25 for HMPOs and 369$\pm$25 for UC HIIs. Therefore, although the
HMSCs have the highest $^{15}$N-enrichment, the mean values for the three 
evolutionary categories are consistent within the errors, which indicates that time
does not seem to play a role in the fractionation of nitrogen (at least until the
formation of a \HII\ region). 
In the top panels of Fig.~\ref{15-NAndamenti} the $^{14}$N/$^{15}$N ratios 
calculated for HCN in the 27 sources are shown as a function of the 
Galactocentric distance, of the line width, and of the kinetic temperature, 
respectively: again, there is no evidence of a trend of these ratios with any 
of the parameters adopted. In particular, the lack of correlation with both temperature and line width, thought to increase both with the evolution of the
source, confirms the independence between the $^{14}$N/$^{15}$N ratio with
the core age. The lack of correlation with evolutionary parameters was found 
also by Fontani et al.~(\citeyear{fontani15a}) in N$_2$H$^+$, but the
dispersion of the ratios is clearly much smaller in our study (from 180 up to 1300 in 
Fontani et al.~\citeyear{fontani15a} and from 250 up to 650 in this work). This may also be due to the fact that they have on average larger uncertainties.

Let us examine now the $^{15}$N- fractionation found for HNC: also here 
we do not find a large spread of values, if compared with that found in Fontani et 
al.~(\citeyear{fontani15a}). In fact, the $^{14}$N/$^{15}$N ratios 
are distributed within $\sim$ 250 and 630 (top-right panel in Fig.~\ref{15-N}).
The mean values for the three evolutionary stages are: 428$\pm$40 for HMSCs, 
462$\pm$31 for HMPOs and 428$\pm$29 for UC HIIs, namely consistent 
within the errors with the $^{14}$N/$^{15}$N ratio measured for the PSN 
of about 441 (from the Solar wind). Also for this molecule it was not found a trend between the isotopic ratio and either the line width or the kinetic temperature 
(bottom panels of Fig.~\ref{15-NAndamenti}).
In conclusion, for both ratios, time does not seem to play a role. 

The lack of correlation between the $^{14}$N/$^{15}$N isotopic ratios and evolutionary parameters or physical parameters (FWHM, T$_{k}$) are somewhat consistent with the chemical model of Roueff et al.~(\citeyear{roueff}) who predict no fractionation of HCN and HNC in cold and dense conditions. However, their models are more appropriate to low-mass dense cores, with a T$_{k}$ of 10 K, than to the warmer high-mass prestellar objects of the type studied in this work. This may explain the lack of carbon fractionation in our work, in constrast to their predictions.
In fact we stress that $^{13}$C in theory can suffer for possible reduced abundances 
due to the fact that nitriles and isonitriles are predicted to be significantly 
depleted in $^{13}$C (Roueff et al.~\citeyear{roueff}). However, this
depletion factor is at most a factor 2 (see Fig.~4 in Roueff et 
al.~\citeyear{roueff}) and is derived from a chemical model with fixed
kinetic temperature of 10~K, which is not the average kinetic temperature
of our sources (not even the starless cores, see Table \ref{risultati1}). Therefore, the
predictions of this model may not be appropriate for our objects. Furthermore, the $^{13}$C-fractionation is dependent on the time and temperature evolution (Szűcs et al.~2014, Röllig \& Ossenkopf~\citeyear{rollig}) and observational tests to check whether this theoretical effect is real have yet to 
be performed.

Intriguingly, doing the Kolmogorov-Smirnov test, a non-parametric statistical 
test, to check if the two data sets (the first are the HC$^{14}$N/HC$^{15}$N 
ratios and the second are the H$^{14}$NC/H$^{15}$NC ratios) 
belong to the same distributions, we found a 
P-value=0.078. The test indicates independence only if P $\leq$ 0.05, and this could mean that with our result we can not conclude anything about the dependence or the independence of the two samples. What is known from the literature is that H$^{15}$NC and HC$^{15}$N can 
form through the same reactions 
(see Terzieva \& Herbst~\citeyear{terzieva} and Wirstr\"{o}m et al.~\citeyear{wirstrom}).
We also have checked the H$^{15}$NC/HC$^{15}$N column density ratio and we 
found that this is roughly more than 1 for HMSCs (with an average value of 
$\sim1\pm0.5$) and roughly less than 1 for HMPOs and UC HIIs (with an average 
value of $\sim0.3\pm0.1$). Loison et al.~(\citeyear{loison}) show that in cold 
regions, where C and CO are depleted, the dissociative recombination of 
HCNH$^{+}$ acts to isomerize HCN into HNC, producing HCN/HNC ratios 
close to or slightly above one. In hot gas, the CN + H$_{2}$ reaction with large 
activation energy (2370 K; Jacobs et al.~\citeyear{jacobs}) and the isomeration 
process C + HNC $\rightarrow$ C + HCN (Loison et al.~\citeyear{loison}) may be responsible 
for the large HCN/HNC measured in hot cores (Schilke et al.~\citeyear{schilke}) 
and in Young Stellar Objects (Schöier et al.~\citeyear{schoier}). The fact that our 
HC$^{15}$N/H$^{15}$NC ratios are always close to one appears to imply that both 
species are probing similar low temperature gas. 



Interestingly, the $^{14}$N/$^{15}$N ratio measured towards the source 
HMSC-G034F2 is an outlier (too high), for the distribution of the ratios in both
HCN and HNC ( but with large errors compared to other objects): a similar high value was found by Bizzocchi et al.~(\citeyear{bizzocchi}) 
in L1544 that is a typical low-mass pre--stellar core. In particular, Bizzocchi et al. 
found a value of $^{14}$N/$^{15}$N$\simeq$1000$\pm$200 from the N$_{2}$H$^{+}$. 
This result, and those measured in HMSC-G034F2, are not consistent with the predictions 
of current models on nitrogen chemistry for N$_{2}$H$^{+}$, HCN and HNC (Roueff et al.~\citeyear{roueff}, Hily-Blant et al.~\citeyear{hily-blant13a}). 
The $^{14}$N/$^{15}$N ratios in HCN and HNC for cores G028-C1, G034-C3, G028-F2, G034-F1, G034-G2 have also been studied indipendently by Zeng et al.~( \citeyear{zeng}) and both works obtain consistent results (within a factor of two).  Taking into account HC$^{15}$N in G028-C1 and comparing our spectrum with that of Zeng et al.~(\citeyear{zeng}) at the same velocity resolution of 0.68 km/s we have obtained a r.m.s of 0.01 K, while they have obtained a r.m.s of 0.02 K. Our T$_{MB}^{peak}$ is 0.03 K that is more than 2.5 $\sigma$, so we can refer to this line as a tentative detection, while Zeng et al.~(\citeyear{zeng}) have a r.m.s that is comparable with the peak of our line. They have also derived the upper limit of the column density using a $\Delta v_{1/2}$ of the line of 2 km/s, while with our fit we obtained a $\Delta v_{1/2}$  of 4 km/s: this is the reason for the factor 2 of difference in the integrated area of this line in the two works. This effect is the same in the other spectra (please compare Fig. A1, A2 and A3 of their work with our Fig.~B1, B2, B3 and B4), and the different integrated areas are also due to the different resolution and sensitivity.

\subsection{D-Fractionation}
\begin{figure}[htpb!]
\includegraphics[width=40pc]{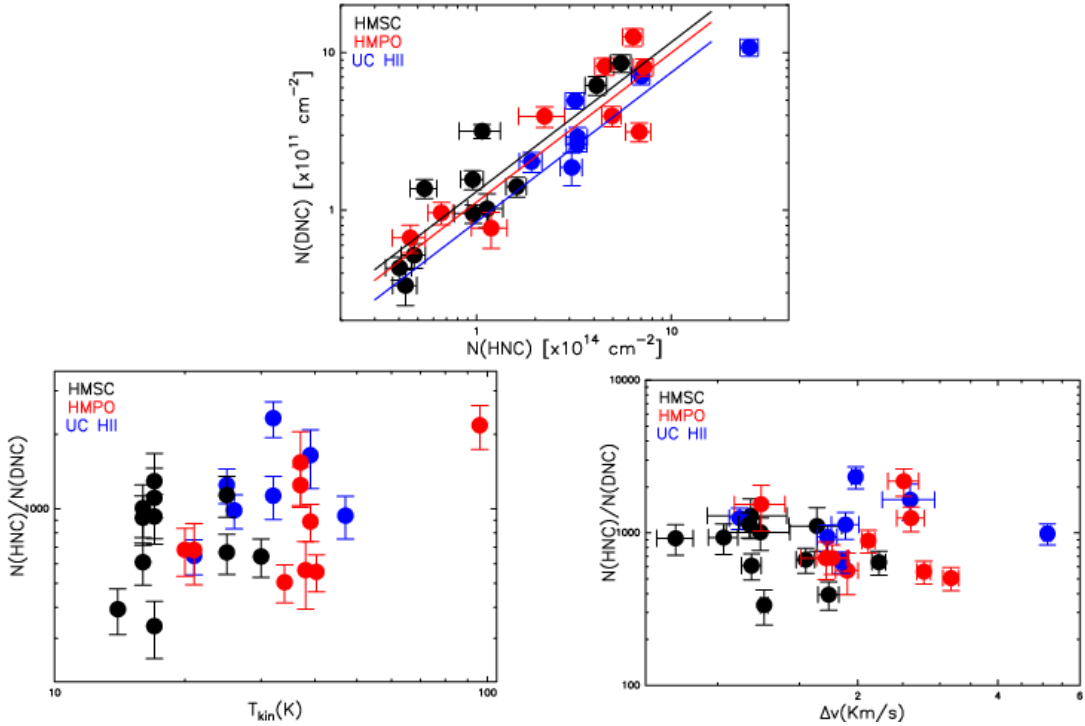}
\caption{Top panel: column density of HNC against that of DNC. The three lines represent the mean values of D/H (see text) for different evolutionary stages: in
black for HMSCs, in red for HMPOs and in blue for UC HIIs. Bottom panels: N(HNC)/N(DNC) as a function
of line widths and kinetic temperatures.
The symbolism of filled circles is the same used for Fig.\ref{15-N}.}
\centering
\label{Deuterium}
\end{figure}

In the top panel of Fig.~\ref{Deuterium} the comparison between the column densities of DNC and HNC,
derived as explained in Sect.~\ref{res}, is shown. The N(HNC)/N(DNC) mean values obtained for the three evolutionary stages are: 823$\pm$94 for HMSCs, 983$\pm$189 for HMPOs and 1275$\pm$210 for UC HIIs, indicated
by three different lines in Fig.~\ref{Deuterium}. Therefore, the D/H mean ratios are: (1.4$\pm$0.2)$\times10^{-3}$ for HMSCs, (1.3$\pm$0.2)$\times10^{-3}$ for HMPOs and 
(0.9$\pm$0.1)$\times10^{-3}$ for UC HIIs.  The D/H mean values were obtained taking all the D/H values (Tab.~\ref{risultati1}) and compute for them the average.
 We note that the D/H values are slightly higher in the early 
stages, but due to the large dispersions, the differences between the three evolutionary
categories are not statistically significant. This result confirms the marginally decreasing trend found by Fontani et al.~(\citeyear{fontani14}), derived from the DNC (1--0) transition in a subsample of
the sources observed in this work. Fontani et
al.~(\citeyear{fontani14}) found average D/H of 0.012, 0.009 and 0.008 in HMSCs, HMPOs and UC HIIs, respectively, with no statistically significant differences among the three evolutionary groups.

\subsection{Comparison between D/H and $^{14}$N/$^{15}$N ratios}
Considering both D- and N-fractionation for HNC toward the same sources, 
we can find a hint whether the two fractionations are linked. The bottom-right 
panel in Fig.~\ref{15-N} show HNC/DNC as a function of HNC/H$^{15}$NC. 
The data show an independence between the two data sets and this can be shown computing the Kendall $\tau$ test. This is a non-parametric test used to measure the ordinal association between two data sets; its definition is:
\[
\tau=\frac{(\text{$\#$ concordant pairs})-(\text{$\#$ discordant pairs})}{\frac{n(n-1)}{2}},
\]
 with $n$ the number of the total pairs.
If $\tau=1$ there is a full correlation, if $\tau=-1$ there is full anticorrelation and if $\tau=0$ there is independence between the two data sets.  We have chosen this statistical test because, compared to other non-parametric tests (i.e. the Spearman's $\rho$ correlation coefficient) this is more robust; moreover, this allows us to compare our analysis with that performed by Fontani et al.~(\citeyear{fontani15a}), who used the same test. 
Following this, we have computed $\tau$ for the two data sets: 
HNC/DNC and HNC/H$^{15}$NC and we have found $\tau\sim0.13$, 
thus independence between D- and $^{15}$N- fractionation in these sources for HNC. This result 
must be compared with that of Fontani et al.~(\citeyear{fontani15a}) that suggested a possible
anti-correlation between the two isotopic ratios for N$_{2}$H$^{+}$. This finding
arises mostly from the fact that the $^{14}$N/$^{15}$N ratio does not vary with
the core evolution, while the D-fractionation shows a faint decreasing trend. This
result indicates that the parameters that cause D-enrichment in HCN and HNC may not influence the fractionation of nitrogen. The independence that 
we have found reflects what is found in some pristine 
Solar System material, in which the spots of high D-enrichments are not always
spatially coincident with those with high $^{15}$N enrichment. 
From the point of view of the models, as discussed in Wirstr\"{o}m 
et al.~(\citeyear{wirstrom}), the D- and $^{15}$N- enrichments do not need to be spatially 
correlated (although could be produced by the same mechanism, 
 i.e. exothermic reactions due to the different zero point energy of the heavier isotope), 
because relevant reactions for D- and $^{15}$N- enrichments have different energy barriers. 
The recent model of Roueff et al.~(\citeyear{roueff}) have reviewed some of the 
reactions in $^{15}$N- fractionation, and concluded that other modeling work is necessary 
to fully understand the relation of the two fractionation
processes.

\subsection{Comparison between N(HNC) and N(HCN)}
\begin{figure}[htpb!]
\centering
\includegraphics[width=30pc]{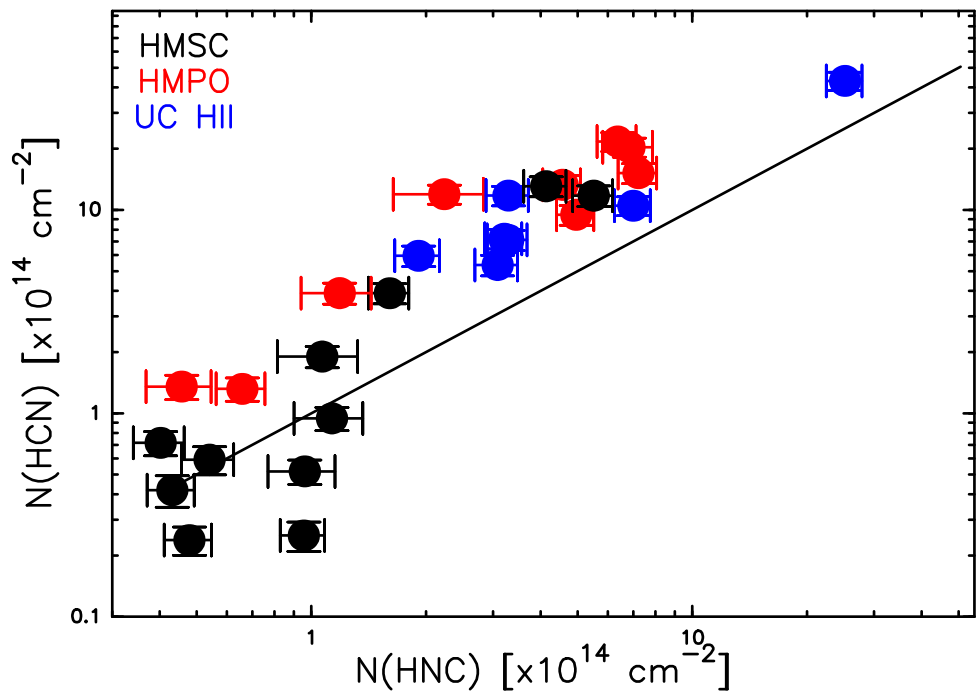}
\caption{Column density of HCN against that of HNC. The solid line is the locus where N(HCN)/N(HNC) is equal to 1 as statistically expected.
The symbolism of filled circles is the same used for Fig.\ref{15-N}.}
\centering
\label{HNC-HCN}
\end{figure}

For completeness we have calculated also the ratio between N(HNC) and N(HCN).
In Fig.~\ref{HNC-HCN} it is shown the column density of HCN versus the HNC 
one, and we note immediately that HMSCs show values of HCN/HNC$\lesssim$1 unlike 
HMPOs and UC HIIs, for which the ratio is $>$1. Mean values are 0.8 $\pm$0.2 for HMSCs, 
2.5 $\pm$0.2 for HMPOs and 2.0$\pm$0.2 for UC HIIs.

We have computed the  HNC/HCN ratio in these two data sets: HMSCs and HMPOs+UC HIIs and using the non-parametric statistical test Kolmogorov-Smirnov
we obtain a P-value=0.025, which indicates that the two distributions are different.
The work of Schilke et al.~(\citeyear{schilke}) shows predictions and observations 
for the Orion hot core OMC-1,  that is the prototype of a high-mass hot core. They found that the HCN/HNC abudance ratio 
is very high ($\sim$80) in the vicinity of Orion-KL, but it declines rapidly in 
adjacent positions to values of $\sim$5. They compared the observations with 
the predictions of the molecular cloud chemistry and found agreement with 
steady state models. More recently Jin et al.~(\citeyear{jin}) found that the abundance 
ratio increases if the sources evolved from IRDCs to UC HIIs and they
suggested that this can happen for neutral-neutral reactions where HNC is 
selectively consumed for T$\gtrsim$24 K (Hirota et al.~\citeyear{hirota}):
\[
 {\rm HNC+H} \rightarrow {\rm HCN + H};
 \]
 \[
 {\rm HNC+O} \rightarrow {\rm NH +CO}.
\]
For the first reaction it was found a 2000 K barrier and for the second there is not yet been theoretical and experimental studies but if it is possible N(HNC)/N(HCN) would decrease rapidly in warm gas. Therefore we emphasize the importance of future laboratory measurements of these latest reactions, especially at low temperatures.

\subsection{Comparison with the $^{14}$N/$^{15}$N galactocentric gradient }
\begin{figure}[htpb!]
\centering
\includegraphics[width=30pc]{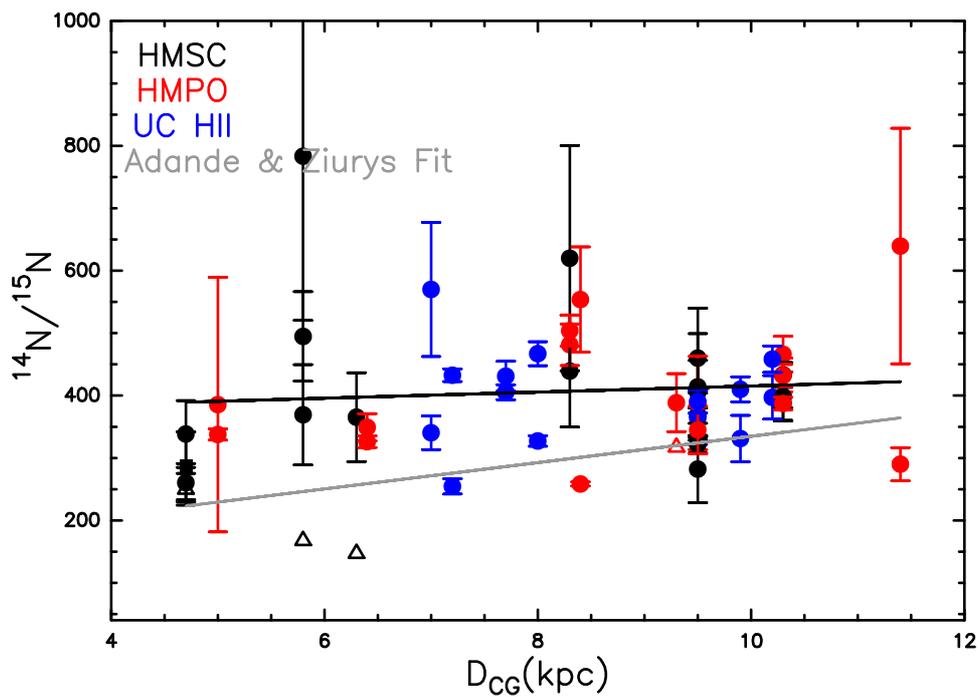}
\caption{ $^{14}$N/$^{15}$N ratios as a function of Galactocentric distances (for HNC and HCN together). The symbolism is the same used for bottom panels of Fig.\ref{15-N}. The black line is the linear fit computed for the plotted data and the grey line is the Adande \& Ziurys (\citeyear{adande}) fit. }
\label{gradiente}
\end{figure}

\begin{figure}[htbp!]
\centering
\includegraphics[width=30pc]{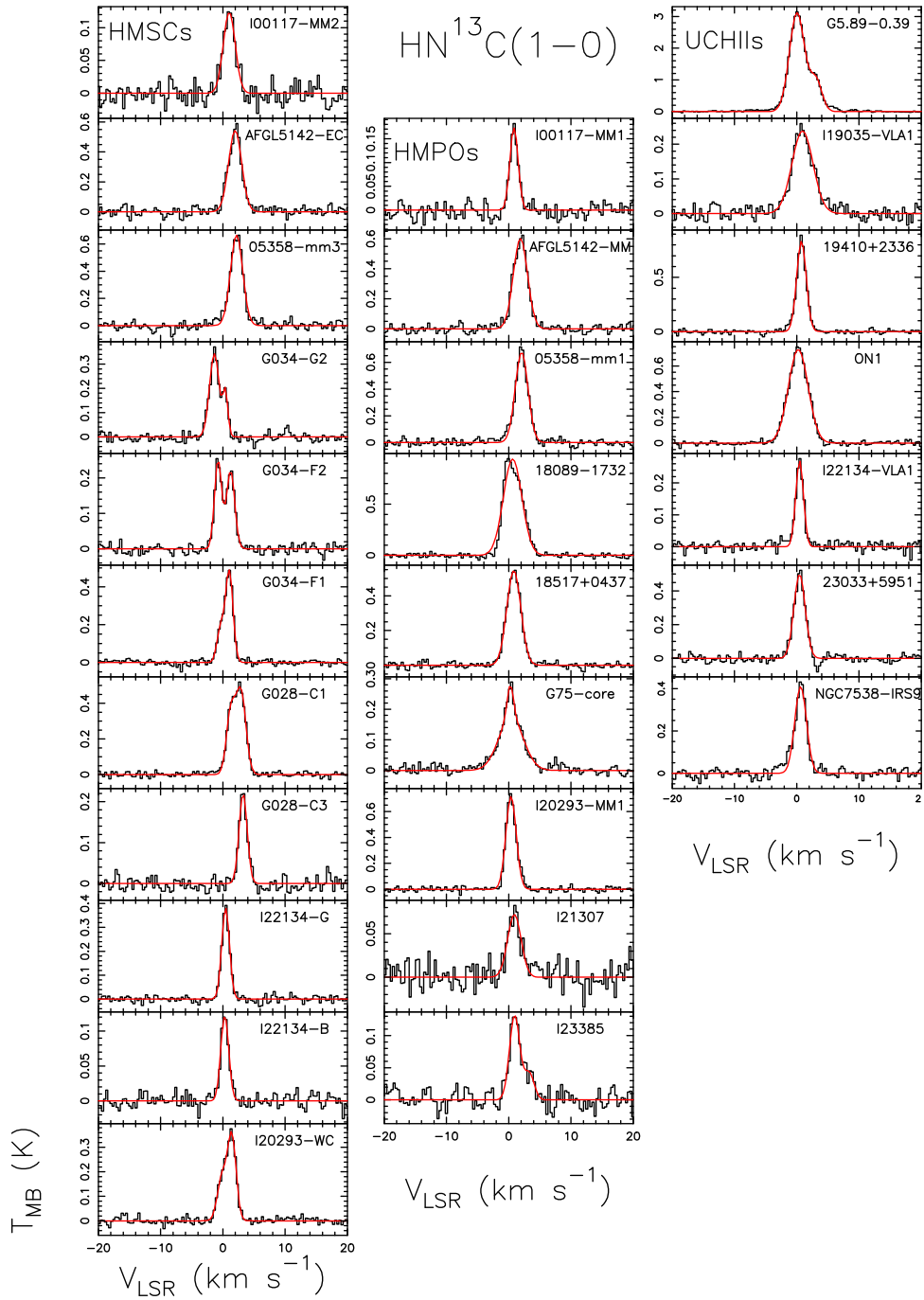}
\caption{Spectra of HN$^{13}$C(1-0) obtained for the sources classified as HMSCs (first column), HMPOs (second column) and UC HII regions (third column) . For each spectrum the x-axis represents a velocity interval of $\pm20$ km s$^{-1}$ around the systemic velocity listed in Tab.~1 of Fontani et al.~(\citeyear{fontani15b}). The y-axis shows the intensity in main beam temperature units. The red curves are the best Gaussian fits obtained with CLASS. Note that for some sources (I20293-WC, G034-G2, G034-F2, G5.89-0.39, G034-F1, G028-C1 and I23385) we have observed two components: we have fitted both lines, and we have used only the one centered on the systemic velocity of the source to compute the column densities. }
\centering
\label{hn13c}
\end{figure}

\begin{figure}[htbp!]
\centering
\includegraphics[width=30pc]{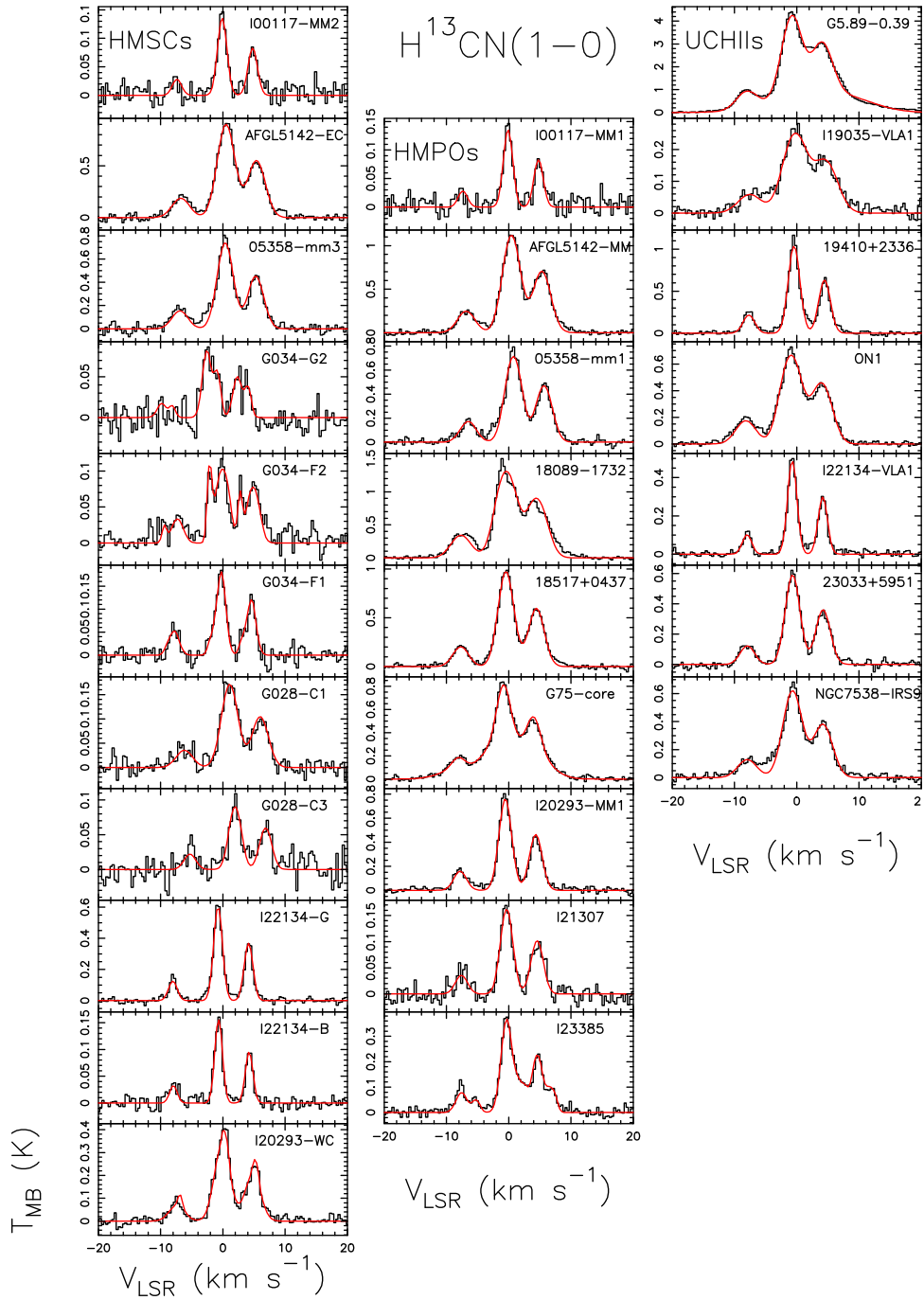}
\caption{Same as Fig.\ref{h13cn-fit} for H$^{13}$CN(1-0). Here the red curves are the best hyperfine fits obtained with CLASS. Please note the presence of the second velocity component in the same sources indicated in the caption of Fig.\ref{hn13c}.}
\label{h13cn}
\end{figure}

\begin{figure}[htpb!]
\centering
\includegraphics[width=30pc]{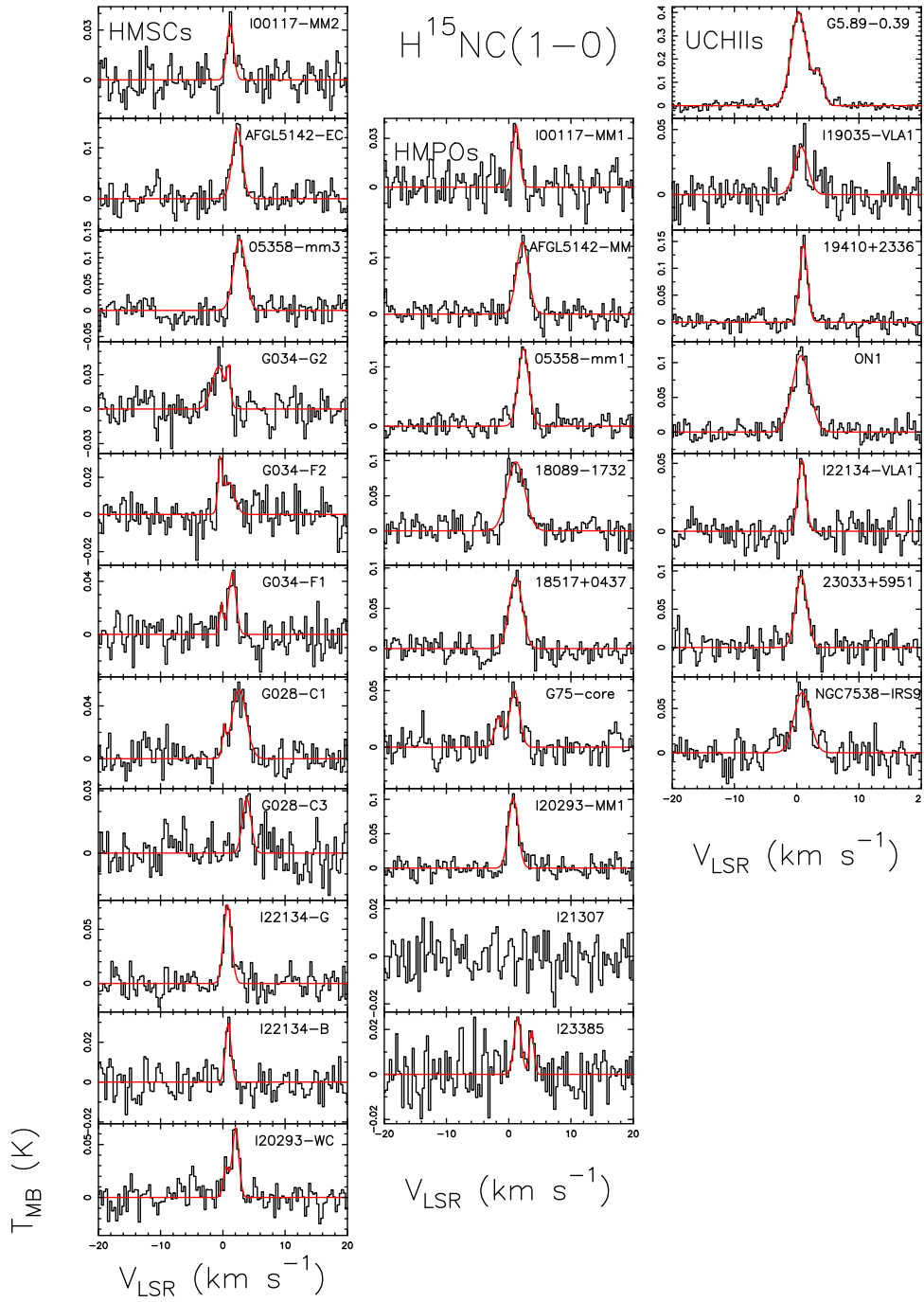}
\caption{ Same as Fig.\ref{h13cn-fit} for H$^{15}$NC(1-0). For I21307 this line was not detected and we have obtained a column density upper limit, as explained in Sect.~3.1.}
\label{h15nc}
\end{figure}

\begin{figure}[htpb!]
\centering
\includegraphics[width=30pc]{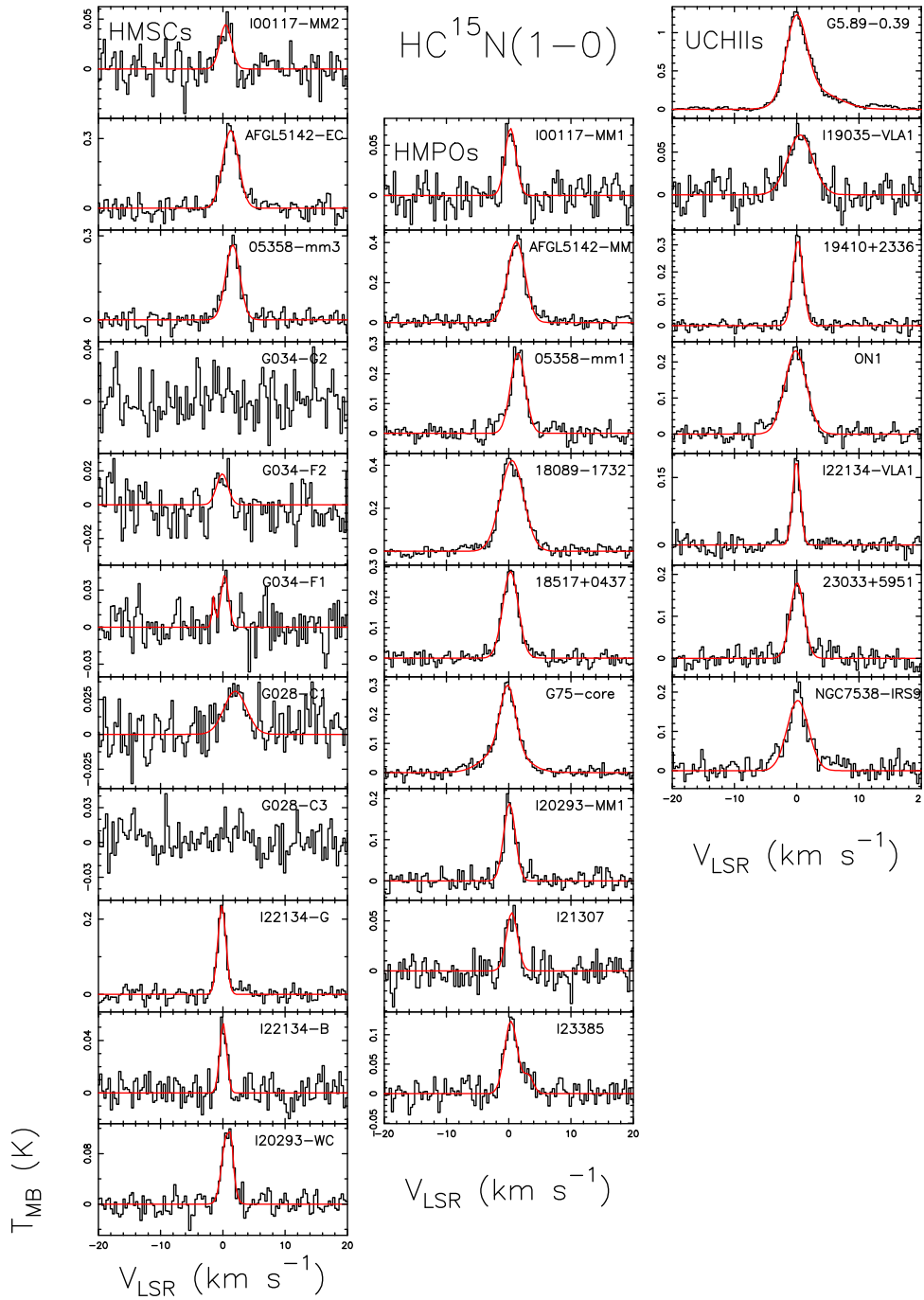}
\caption{ Same as Fig.\ref{h13cn-fit} for HC$^{15}$N(1-0). For G034-G2, G028-C3 and G034-F2  the line was not detected and we obtained a column density upper limit, as explained in Sect.~3.1. For G034-F1 we did not have a clear detection but we have obtained from the Gaussian fit that $T_{MB}^{peak}\gtrsim2.5\sigma$, we have computed as usual the column density and we refer to this latter as "tentative detection" (see also Table \ref{hc15n-dnc-fit}).}
\label{hc15n}
\end{figure}

\begin{figure}[htpb!]
\centering
\includegraphics[width=30pc]{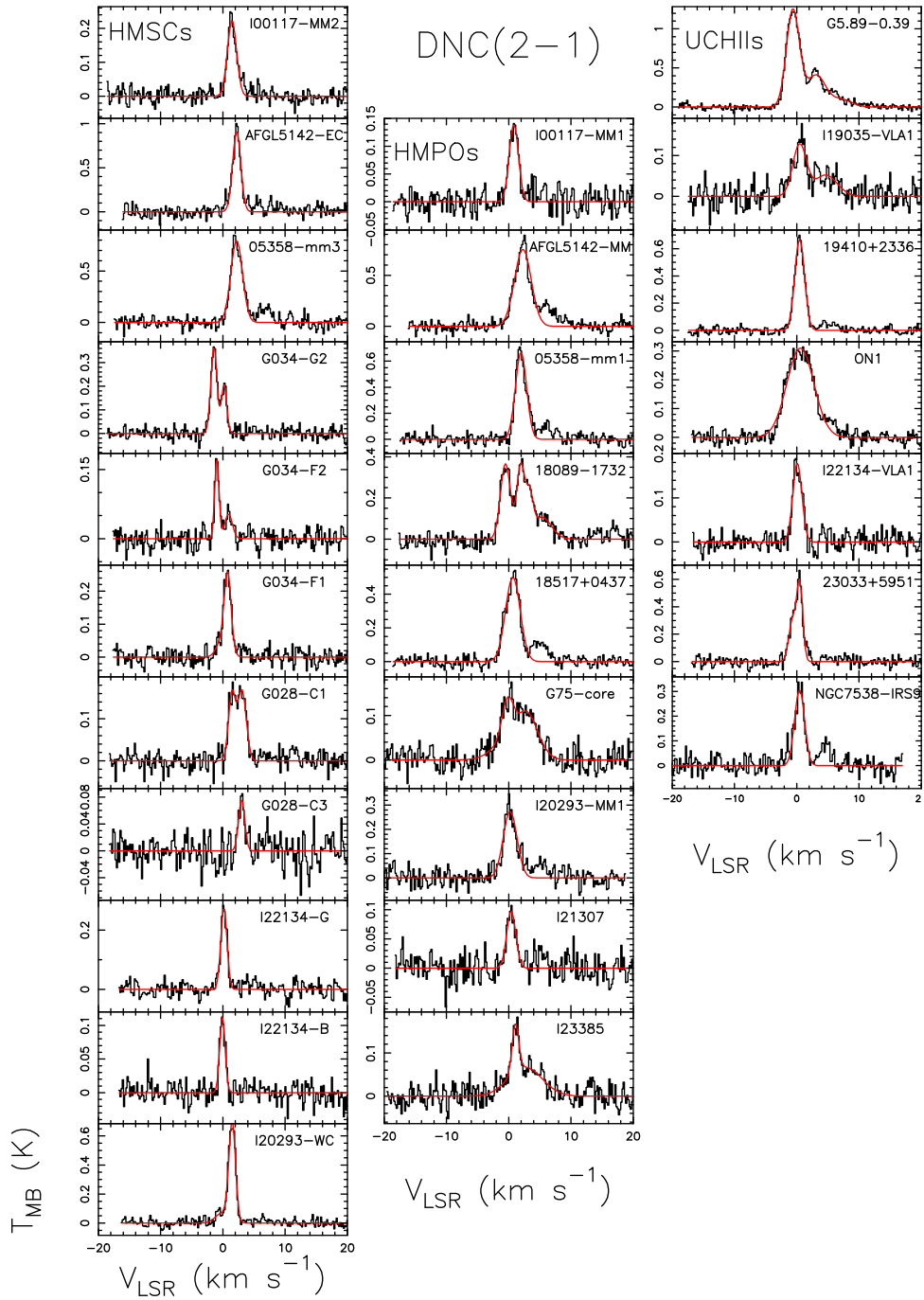}
\caption{ Same as Fig.\ref{h13cn-fit} for DNC(2-1). For evolved sources (HMPOs and UC HII) there is also the acetaldehyde line as explained in Sect.~3.2: when it was possible we have excluded the line from the Gaussian fit, otherwise we have fitted the two lines together and used only the DNC(2-1) to compute the column densities.}
\label{dnc}
\end{figure}

Adande \& Ziurys (\citeyear{adande}) evaluated the $^{14}$N/$^{15}$N ratio across the galaxy through millimeter observations of CN and HNC in regions of star formation. They have enlarged the sample also with the high-mass sources observed by Dahmen et al.~(\citeyear{dahmen}) via HCN observations. In order to compare with their work we have plotted our $^{14}$N/$^{15}$N ratios (for HCN and HNC together) as a function of the galactocentric distance of the sources (see Table \ref{alldata} and Fig.~\ref{gradiente}). We have computed a linear fit and we have obtained the relation:

\begin{equation}
^{14}\text{N}/^{15}\text{N}=(4.9\pm7.6) \text{ kpc}^{-1} \times D_{GC} + (366.5\pm65.6)
\end{equation}
that is not in agreement with that obtained by Adande \& Ziurys (\citeyear{adande}):
\begin{equation}
^{14}\text{N}/^{15}\text{N}=(21.1\pm5.2) \text{ kpc}^{-1} \times D_{GC} + (123.8\pm37.1).
\end{equation}
This indicates that in our dataset, the gradient found by Adande \& Ziurys (\citeyear{adande}) is not confirmed. We speculate that, as the beam used by Adande \& Ziurys (\citeyear{adande}) is larger, and hence more influenced by the diffuse surrounding material, the difference in ratios may be due to the material that is around the target cores, in which this ratio may be lower.
We note also that Adande \& Ziurys (\citeyear{adande}) derived their fit using the source SgrB2(NW),
located at a galactocentric radius of 0.1 kpc: for this source, they obtained a $^{14}\text{N}/^{15}$N lower limit of 164.
This value strongly influences their fit results, and hence could have a strong impact on both
the slope and the intercept with the y-axis. Therefore, any comparison between the two fit results
must be taken with caution.

\section{Conclusions}
We have observed the \emph{J}=1-0 rotational transitions of H$^{15}$NC, HN$^{13}$C, H$^{13}$CN and HC$^{15}$N, together with \emph{J}=2-1 transition of DNC, towards 27 massive star forming cores in different evolutionary stages, in order to derive the $^{15}$N- and D-fractionation, and to compare each other. We find $^{14}$N/$^{15}$N in HCN between $\sim$200 and $\sim$700, and in HNC between $\sim$260 and $\sim$600, with a small spread around the PSN value of 441. Comparing the $^{14}$N/$^{15}$N ratios for different evolutionary stages we do not find any trend, indicating that time does not seem to play a role in the N-fractionation;  furthermore we can not say anything about the correlation between the $^{15}$N-fractionation for the two molecules, HNC and HCN. Considering both D- and $^{15}$N-fractionation for HNC toward the same sources, we find no correlation. This is consistent with the lack of correlation found by Fontani et al.~(\citeyear{fontani15a}) in N$_{2}$H$^{+}$: the causes of D-enrichment in HCN and HNC do not affect the $^{15}$N-fractionation. Our findings are in agreement with the recent chemical models of Roueff et al.~(\citeyear{roueff}). 
The independence between D/H and $^{14}$N/$^{15}$N ratios confirms the recent findings of Guzm\'{a}n et al.~(\citeyear{guzman}) in protoplanetary disks. Conversely, the low values of $^{14}$N/$^{15}$N found by them (and the tentative decrease with decreasing distance to the
star) is not in contrast with our findings, because our results are obtained on angular scales much larger than that of a protoplanetary disk. All this indicates that the $^{15}$N enrichment is a local effect, which does not involve the larger-scale envelope.\\\\
\emph{Acknowledgments.} We are grateful to the IRAM-30m telescope staff for their help during the observations. 
Many thanks to the anonymous Referee for the careful reading of the paper and his/her comments that improved the work. 
Paola Caselli and Luca Bizzocchi acknowledge support from the European Research Council (project PALs 320620).


{}
\onecolumn

\begin{table*}
\normalsize
\begin{center}
 \caption{Total column densities (beam-averaged), computed as explained in Sect.~3, of H$^{15}$NC, HN$^{13}$C and DNC. In the third and fifth columns there are the error on column densities without considering the calibration error ($\Delta$N). In the
 seventh, eighth and ninth columns there are the corresponding $^{14}$N/$^{15}$N and D/H isotopic ratios. Uncertainties in the column densities and in the isotope ratios have been computed as explained in Sect.~3.1 and 3.2. In the last column there are the kinetic temperatures of the clumps derived from Fontani et al.~(\citeyear{fontani15a}): for the sources without a derivation of T$_{k}$, the mean value for that evolutionary stage was taken (for the HMSCs the average was done without the "warm" ones, i.e those with a T$_{k}<20$ K).}
\label{risultati1}
\begin{tabular}{llllllllll}
 \hline \hline
 Source & N(H$^{15}$NC)& $\Delta$N& N(HN$^{13}$C)& $\Delta$N& N(DNC)& $\frac{HNC}{H^{15}NC}^{*}$ & $\frac{HNC}{DNC}^{**}$ & $\frac{DNC}{HNC}$  & T$_{k}$\\
          & \multicolumn{2}{c}{($\times10^{10}$cm$^{-2}$) }  & \multicolumn{2}{c}{($\times10^{11}$cm$^{-2}$) } & ($10^{11}$cm$^{-2}$) & &  &  & (K)\\
 \cline{1-10}
 \multicolumn{10}{c}{HMSC} \\
 \cline{1-10}
  I00117-MM2 & $12\pm4$ &  2&  $8\pm1$  & 0.4&$4.2\pm0.6$ & $460\pm80$& $406\pm 77$ & $ (2.5\pm0.5)\times 10^{-3}$& $14$\\
  AFGL5142-EC$^{w}$ & $104\pm20$ &  10 &  $56\pm7$ & 1&$19\pm3$ &$398\pm39$ &$674\pm136$ & $(1.5\pm0.3)\times 10^{-3}$& $25$ \\
  05358-mm3$^{w}$ & $141\pm24$ & 10&  $74\pm9$  &1&$27\pm3$ & $388\pm28$&$627\pm103$&$(1.6\pm0.3)\times 10^{-3}$ & $30$\\
  G034-G2(MM2)  & $26\pm8$ &  5 &  $19\pm3$ & 0.6&$4.8\pm0.7$ &$365\pm71$ & $612\pm131$&$(1.6\pm0.3)\times 10^{-3}$ & $16$*\\
  G034-F1(MM8)  & $19\pm4$ &2 &  $20\pm4$ &2 &$2.9\pm0.4$ &$495\pm72$ & $1002\pm243$& $ (1.0\pm0.2)\times 10^{-3}$& $16$* \\
  G034-F2(MM7)  & $6\pm3$ &  2& $10\pm1$ &0.4 &$1.6\pm0.3$ & $783\pm263$& $908\pm193$& $(1.1\pm0.2)\times 10^{-3}$& $16$*\\
  G028-C1(MM9) & $43\pm8$ &4&  $28\pm6$ & 2 &$3.2\pm0.8$  &$260\pm30$ & $1081\pm356$& $(0.9\pm0.3)\times 10^{-3}$& $17$ \\
  G028-C3(MM11) & $13\pm4$ & 2  &  $11\pm1$  & 0.4 &$1.0\pm0.3$ &$338\pm53$ &$1360\pm426$ & $(0.7\pm0.2)\times 10^{-3}$& $17$\\
  I20293-WC & $24\pm7$ &  4&  $17\pm4$  & 2 &$10\pm1$ &$439\pm90$ &$326\pm 83$ &$(3.1\pm0.8)\times 10^{-3}$& $17$ \\
  I22134-G$^{w}$  & $43\pm8$ &  4&  $23\pm3$ &0.5 & $4.4\pm0.6$&$369\pm35$ & $1114\pm210$ & $(0.9\pm0.2)\times 10^{-3}$& $25$\\
  I22134-B  & $10\pm3$  & 2&  $6.0\pm0.9$  & 0.3&$1.3\pm0.2$ &$414\pm85$ & $984\pm 211$&$(1.0\pm0.2)\times 10^{-3}$ & $17$\\
  \cline{1-10}      
\multicolumn{10}{c}{HMPO} \\
\cline{1-10}
  I00117-MM1 & $16\pm4$ & 3 &  $9\pm1$  &  0.4&$3.0\pm0.5$ &$388\pm75$ & $640\pm128$&$(1.6\pm0.3)\times 10^{-3}$ & $20$\\
  AFGL5142-MM & $147\pm24$ & 9 &  $86\pm10$ &  1&$39\pm5$ & $433\pm27$& $504\pm87$& $(2.0\pm0.3)\times 10^{-3}$& $34$\\
  05358-mm1 & $154\pm25$ &  9&  $97\pm11$  & 2 &$25\pm3$ &$466\pm29$ & $887\pm146$& $ (1.1\pm0.2)\times 10^{-3}$& $39$\\
  18089-1732  & $58\pm34$  & 29 &  $52\pm14$& 9 &$12\pm2$ & $385\pm204$& $576\pm182$ & $(1.7\pm0.5)\times 10^{-3}$& $38$ \\
  18517+0437  & $130\pm21$  &  8&  $89\pm10$ & 1 &$25\pm3$ &$349\pm21$ &$561\pm92$ & $(1.8\pm0.3)\times 10^{-3}$& $40$*\\
  G75-core& $124\pm31$   & 18&  $109\pm16$ &  5 &$10\pm1$ & $554\pm84$& $2122\pm377$& $(5.0\pm0.9)\times 10^{-4}$& $96$\\
  I20293-MM1  & $103\pm17$ & 7 &  $80\pm9$& 1&$12\pm2$ & $481\pm33$& $1277\pm257$& $(0.8\pm0.1)\times 10^{-3}$& $36$\\
  I21307 & $\leq 15^{u}$  & -- &  $7\pm1$ & 0.6&$2.1\pm0.4$ & $\geq317$& $700\pm167$&$ (1.4\pm0.4)\times 10^{-3}$ & $21$\\
  I23385  & $19\pm7$ &  5 &$15\pm3$&  2&$2.4\pm0.6$ & $639\pm189$& $1564\pm 501$&$(0.6\pm0.2)\times 10^{-3}$ & $37$\\
    \cline{1-10}													      	  
\multicolumn{10}{c}{UC HII} \\
\cline{1-10}
  G5.89-0.39  & $576\pm70$  &13 &  $453\pm49$ & 3&$33\pm4$ & $432\pm10$& $2333\pm379$ & $ (4.3\pm0.7)\times 10^{-4}$ & $32$*\\
  I19035-VLA1 & $54\pm16$ & 10&  $57\pm7$ &  2&$6\pm1$ &$570\pm107$ & $1585\pm328$& $(0.6\pm0.1)\times 10^{-3}$& $39$\\
  19410+2336 & $74\pm11$ & 4&  $55\pm6$& 0.6 &$15\pm2$ &$431\pm24$ & $657\pm113$&$(1.5\pm0.3)\times 10^{-3}$ & $21$\\
  ON1  & $149\pm21$ &  6&  $116\pm12$&  0.9& $22\pm2$& $467\pm19$& $978\pm 135$& $ (1.0\pm0.1)\times 10^{-3}$& $26$\\
  I22134-VLA1& $53\pm11$ &6 &  $28\pm4$ & 0.9&$6.3\pm0.9$ &$364\pm43$ & $948\pm191$& $(1.1\pm0.2)\times 10^{-3}$& $47$\\
  23033+5951& $82\pm15$ & 7&  $44\pm5$ &0.9& $8.1\pm0.9$&$ 397\pm35$ & $1242\pm197$ & $(0.8\pm0.1)\times 10^{-3}$& $25$\\
  NGC 7538-IRS9 & $100\pm21$  &11 & $46\pm6$ &1 & $9\pm1$&$331\pm37$ & $1137\pm195$&$(0.9\pm0.1)\times 10^{-3}$ & $32$*\\
\hline
\end{tabular}
\end{center}
$^{*}$ it has been multiplied by $\frac{^{12}C}{^{13}C}$ as described in Section (3.1) and given in Table \ref{comparisondifferentmolecules};

$^{**}$ it has been multiplied by $\frac{^{12}C}{^{13}C}$ and by the correction of the different beams 3.09;

$^{t}$ tentative detection;

$^{u}$ upper limits;

$^{w}$: ``warm'' HMSC;

* average value for the specific evolutionary stage.
\end{table*}
\normalsize

\begin{table*}
\begin{center}
 \caption{Total column densities (beam-averaged), computed as explained in Sect.~(3), of HC$^{15}$N and H$^{13}$CN(1-0) transitions. In the third and fifth columns there are the error on column densities without considering the calibration error ($\Delta$N). In the
 sixth column there is the corresponding $^{14}$N/$^{15}$N isotopic ratios. Uncertainties have been derived as explained in the caption of Table \ref{risultati1}. In the last column there are the kinetic temperatures of the clumps derived from Fontani et al.~(\citeyear{fontani15a}).}
\label{risultati2}
\begin{tabular}{lllllll}
 \hline \hline
 Source & N(HC$^{15}$N) & $\Delta$N& N(H$^{13}$CN) & $\Delta$N&$\frac{HCN}{HC^{15}N}^{*}$ & T$_{k}$\\
           & \multicolumn{2}{c}{($\times10^{10}$cm$^{-2}$) } & \multicolumn{2}{c}{($\times10^{11}$cm$^{-2}$) } &  & (K) \\
 \cline{1-7}
 \multicolumn{7}{c}{HMSC} \\
 \cline{1-7}
 I00117-MM2 &  $22\pm6$ &  4& $9\pm1$ &0.5  & $ 282\pm54$ & $14$\\
  AFGL5142-EC$^{w}$ &  $327\pm47$ & 14 & $176\pm20$& 2&$398\pm18$ &$25$\\
  05358-mm3$^{w}$  &  $270\pm39$  & 12& $158\pm18$& 2 &$433\pm20$ & $30$\\
  G034-G2(MM2) &  $\leq 17^{u}$  & -- & $5\pm1$ &  0.3&$\geq147$& $16$*  \\
  G034-F1(MM8)  &  $14\pm4$  &  3& $11\pm2$ &  0.4&$369\pm80$ & $16$*\\
  G034-F2(MM7)   &  $\leq 14^{u}$&  -- &$5\pm1$&0.3 &$\geq 168$& $16$* \\
  G028-C1(MM9) &  $32\pm9^{t}$ & 6 & $23\pm3$&  0.7 &$287\pm55$& $17$ \\
  G028-C3(MM11)&  $\leq 16^{u}$ &  --& $10\pm2$& 0.8&$\geq250$ & $17$ \\
  I20293-WC  &  $31\pm12$ & 9& $31\pm4$& 0.6&$620\pm180$ & $17$ \\
  I22134-G$^{w}$   &  $120\pm17$ & 5 & $56\pm6$ &  0.8&$322\pm14$& $25$ \\
  I22134-B   &  $17\pm4$  &  2& $10\pm1$&  0.4&$406\pm50$ & $17$\\
  \cline{1-7}      
\multicolumn{7}{c}{HMPO} \\
\cline{1-7}
   I00117-MM1  &  $38\pm8$  &  4& $19\pm2$& 0.6&$345\pm38$ & $20$\\
 AFGL5142-MM  &  $558\pm70$ &  14 & $292\pm31$& 2 &$387\pm10$& $34$  \\
  05358-mm1   &  $351\pm50$ & 14& $205\pm23$&  2 &$432\pm18$& $39$ \\
  18089-1732   &  $354\pm36$  &  9& $278\pm30$&  2&$338\pm9$ & $38$ \\
  18517+0437 &  $405\pm52$ & 11 & $259\pm28$ &  2&$326\pm9$ & $40$*\\
  G75-core &  $790\pm88$  & 9& $324\pm35$& 2 &$258\pm3$ & $96$\\
  I20293-MM1   &  $187\pm28$ &  9 & $152\pm17$& 2 &$504\pm25$ & $36$\\
  I21307  &  $35\pm7$ & 4 & $20\pm3$& 0.7&$389\pm46$ & $21$\\
  I23385&  $134\pm26$ & 12& $48\pm6$ &  0.9&$290\pm26$ & $37$\\
    \cline{1-7}													      	  
\multicolumn{7}{c}{UC HII} \\
\cline{1-7}
G5.89-0.39  &  $1674\pm246$ &  81& $775\pm79$ & 2&$255\pm12$ & $32$*\\
  I19035-VLA1  &  $157\pm28$  & 12 &$99\pm11$& 2 &$340\pm27$ & $39$\\
  19410+2336 &  $176\pm23$ & 5 & $123\pm13$& 1&$405\pm12$& $21$ \\
  ON1   &  $319\pm40$ & 8 &$174\pm18$&1 &$327\pm8$ & $26$\\
  I22134-VLA1  &  $152\pm22$ & 7& $86\pm10$ & 1&$390\pm18$ & $47$\\
  23033+5951  &  $155\pm23$  &7 & $96\pm11$ & 1&$458\pm21$& $25$ \\
  NGC 7538-IRS9 &  $288\pm43$ & 14 & $164\pm18$ & 1 &$410\pm20$ & $32$*\\
\hline
\end{tabular}
\end{center}
$^{*}$ it has been multiplied by $\frac{^{12}C}{^{13}C}$ as described in Section (3.1) and given in Table \ref{comparisondifferentmolecules};

$^{t}$ tentative detection;

$^{u}$ upper limits;

$^{w}$: ``warm'' HMSC;

* average value for the specific evolutionary stage.
\end{table*}
\normalsize

\quad
\quad


\begin{appendix}
\section{Fit results}
 In this appendix, the results of the fitting procedure to the HN$^{13}$C(1-0), H$^{15}$NC(1-0), HC$^{15}$N(1-0), DNC(2-1) and H$^{13}$CN(1-0) lines of all sources (see Sect. 3.1 and 3.2) are shown.
\clearpage
\begin{table*}
\normalsize
\caption{Values obtained with a Gaussian fit to the HN$^{13}$C and H$^{15}$NC(1-0) lines. The errors are coming from the fitting procedure and are not taking account the calibration error on T$_{MB}$. In the second and in the third columns the centroid velocities and the FWHM are listed. In the fourth column there are the integrated intensities and in the fifth column there is the r.m.s of the \textbf{spectra}. The cases in which the line is not detected, and therefore only column density upper limits could be obtained as explained in Sect.~3.1, are indicated with -- . The source names
 are taken from Fontani et al.~(\citeyear{fontani15b}).}
  \tiny
  \begin{tabular}{l*{8}{c}}
  \hline
  \toprule
  Source & \multicolumn{4}{c}{HN$^{13}$C(1-0)} & \multicolumn{4}{c}{H$^{15}$NC(1-0)}\\
  &$v_{LSR}$ & $\Delta v_{1/2}$ & $\int T_{MB}\,dv$ & $\sigma$ & $v_{LSR}$ & $\Delta v_{1/2}$ & $\int T_{MB}\,dv$ & $\sigma$\\
  &(km/s) & (km/s) & (K km/s) & (mK)   &(km/s) & (km/s) & (K km/s) & (mK)\\
  \midrule
  \multicolumn{9}{c}{HMSC}\\
  \midrule
  I00117-MM2 & $-35.27\pm0.06$ &  $2.30\pm0.02$ &  $0.30\pm0.02$ & $10$& $-35.1\pm0.1$ &  $1.3\pm0.4$ &  $0.05\pm0.01$ & $9$\\
  AFGL5142-EC & $-1.92\pm0.03$ &  $2.47\pm0.08$ &  $1.43\pm0.04$ & $30$& $-1.60\pm0.08$ &  $1.9\pm0.2$ &  $0.28\pm0.03$ & $20$\\
  05358-mm3 & $-15.39\pm0.02$ &  $2.32\pm0.06$ &  $1.65\pm0.03$ & $20$& $-14.92\pm0.08$ &  $2.2\pm0.2$ &  $0.32\pm0.02$ & $20$\\
  G034-G2(MM2) & $42.24\pm0.03$ &  $1.84\pm0.08$ &  $0.67\pm0.02$ & $20$& $42.9\pm0.2$ &  $2.4\pm0.6$ &  $0.1\pm0.02$ & $10$\\
  G034-F1(MM8) & $58.74\pm0.06$ &  $1.40\pm0.09$ &  $0.72\pm0.07$ & $10$& $59.20\pm0.09$ &  $1.4\pm0.2$ &  $0.07\pm0.01$ & $10$\\
  G034-F2(MM7) & $56.93\pm0.03$ &  $1.35\pm0.06$ &  $0.36\pm0.01$ & $10$& $57.3\pm0.1$ &  $0.7\pm0.2$ &  $0.02\pm0.008$ & $7$\\
  G028-C1(MM9) & $81.19\pm0.09$ &  $1.9\pm0.1$ &  $0.95\pm0.1$ & $10$& $80.9\pm0.1$ &  $2.7\pm0.3$ &  $0.15\pm0.01$ & $9$\\
  G028-C3(MM11)& $81.56\pm0.03$ &  $1.55\pm0.08$ &  $0.36\pm0.01$ & $10$& $82.1\pm0.2$ &  $1.6\pm0.3$ &  $0.05\pm0.009$ & $8$\\
  I20293-WC & $7.76\pm0.09$ &  $1.6\pm0.1$ &  $0.58\pm0.08$ & $10$& $8.4\pm0.1$ &  $1.2\pm0.2$ &  $0.09\pm0.01$ & $10$\\
  I22134-G & $-17.85\pm0.01$ &  $1.47\pm0.04$ &  $0.60\pm0.01$ & $10$& $-17.55\pm0.06$ &  $1.5\pm0.2$ &  $0.11\pm0.01$ & $10$\\
  I22134-B & $-17.98\pm0.04$ &  $1.5\pm0.1$ &  $0.20\pm0.01$ & $10$& $-17.43\pm0.09$ &  $1.0\pm0.2$ &  $0.03\pm0.006$ & $7$\\
   \midrule
  \multicolumn{9}{c}{HMPO}\\
  \midrule
   I00117-MM1 & $-35.44\pm0.04$ &  $1.56\pm0.09$ &  $0.29\pm0.01$ & $10$& $-35.1\pm0.1$ &  $1.2\pm0.2$ &  $0.05\pm0.008$ & $9$\\
 AFGL5142-MM & $-2.01\pm0.02$ &  $2.68\pm0.05$ &  $1.72\pm0.03$ & $20$& $-1.70\pm0.07$ &  $2.2\pm0.1$ &  $0.30\pm0.02$ & $10$\\
  05358-mm1 & $-15.53\pm0.02$ &  $2.41\pm0.05$ &  $1.73\pm0.03$ & $20$& $-15.23\pm0.06$ &  $2.0\pm0.1$ &  $0.28\pm0.02$ & $10$\\
  18089-1732 & $32.30\pm0.03$ &  $1.6\pm0.1$ &  $0.9\pm0.2$ & $20$& $32.7\pm0.2$ &  $1.4\pm0.3$ &  $0.11\pm0.05$ & $10$\\
  18517+0437 & $44.52\pm0.02$ &  $2.66\pm0.04$ &  $1.54\pm0.02$ & $10$& $44.84\pm0.08$ &  $2.5\pm0.2$ &  $0.23\pm0.01$ & $10$\\
  G75-core & $0.57\pm0.07$ &  $4.6\pm0.3$ &  $0.86\pm0.04$ & $10$& $1.1\pm0.1$ &  $1.9\pm0.4$ &  $0.10\pm0.01$ & $10$\\
  I20293-MM1 & $6.67\pm0.01$ &  $1.95\pm0.03$ &  $1.49\pm0.02$ & $10$& $7.02\pm0.06$ &  $1.8\pm0.1$ &  $0.20\pm0.01$ & $10$\\
  I21307 & $-45.8\pm0.01$ &  $2.6\pm0.3$ &  $0.20\pm0.02$ & $10$& -- &  -- &  -- & $8$\\
  I23385 & $-49.5\pm0.1$ &  $2.0\pm0.2$ &  $0.27\pm0.03$ & $10$ & $-49.1\pm0.2$ &  $1.4\pm0.5$ &  $0.04\pm0.009$ & $9$\\
    \midrule
  \multicolumn{9}{c}{UC H$_{II}$}\\
  \midrule
   G5.89-0.39 & $9.05\pm0.01$ &  $2.87\pm0.02$ &  $9.54\pm0.07$ & $30$& $9.35\pm0.03$ &  $2.93\pm0.08$ &  $1.26\pm0.03$ & $10$\\
  I19035-VLA1 & $33.33\pm0.05$ &  $4.0\pm0.1$ &  $1.01\pm0.03$ & $20$& $33.1\pm0.2$ &  $2.6\pm0.8$ &  $0.1\pm0.02$ & $10$\\
  19410+2336 & $23.16\pm0.01$ &  $1.81\pm0.03$ &  $1.60\pm0.02$ & $20$& $23.50\pm0.03$ &  $1.45\pm0.09$ &  $0.22\pm0.01$ & $10$\\
  ON1 & $12.25\pm0.01$ &  $3.73\pm0.03$ &  $2.89\pm0.02$ & $10$& $12.70\pm0.06$ &  $3.3\pm0.2$ &  $0.38\pm0.02$ & $10$\\
  I22134-VLA1 & $-17.83\pm0.02$ &  $1.45\pm0.06$ &  $0.45\pm0.01$ & $10$& $-17.48\pm0.08$ &  $1.5\pm0.2$ &  $0.08\pm0.009$ & $8$\\
  23033+5951 & $-52.54\pm0.02$ &  $2.15\pm0.05$ &  $0.14\pm0.02$ & $20$& $-52.27\pm0.09$ &  $2.2\pm0.2$ &  $0.22\pm0.02$ & $10$\\
  NGC 7538-IRS9 & $-56.37\pm0.03$ &  $2.21\pm0.08$ &  $0.97\pm0.03$ & $20$& $-56.1\pm0.1$ &  $3.0\pm0.5$ &  $0.22\pm0.02$ & $10$\\
  \bottomrule
 \end{tabular}
\centering
\label{hn13c-h15nc-fit}
\normalsize
\end{table*}

\begin{table}
\normalsize
\caption{Values obtained with Gaussian fits to the HC$^{15}$N and DNC(2-1) lines. The errors are coming from the fitting procedure and do not take the calibration error on T$_{MB}$ into account. In the second and in the third colums are listed the centroid velocities and the line FWHM, respectively. In the fourth column there are the integrated intensities and in the fifth column there is the r.m.s of the spectra. The cases in which the line is not detected,  and therefore only column density upper limits could be obtained as explained in Sect.~3.1, are indicated with -- .}
\tiny
\begin{center}
  \begin{tabular}{l*{8}{c}}
  \hline
  \toprule
  Source &\multicolumn{4}{c}{HC$^{15}$N(1-0)} & \multicolumn{4}{c}{DNC(2-1)}\\
  & $v_{LSR}$ & $\Delta v_{1/2}$ & $\int T_{MB}\,dv$ & $\sigma$& $v_{LSR}$ & $\Delta v_{1/2}$ & $\int T_{MB}\,dv$ & $\sigma$\\
  &(km/s) & (km/s) & (K km/s) & (mK)&(km/s) & (km/s) & (K km/s) & (mK)\\
  \midrule
  \multicolumn{9}{c}{HMSC}\\
  \midrule
  I00117-MM2 & $-35.8\pm0.2$ &  $2.2\pm0.4$ &  $0.11\pm0.02$ & $10$& $-34.83\pm0.03$ &  $1.73\pm0.09$ &  $0.41\pm0.02$ & $20$\\
  AFGL5142-EC & $-2.64\pm0.07$ &  $3.0\pm0.2$ &  $1.07\pm0.05$ & $30$& $-1.67\pm0.03$ &  $1.55\pm0.07$ &  $1.50\pm0.05$ & $60$\\
  05358-mm3 & $-16.01\pm0.06$ &  $2.7\pm0.1$ &  $0.76\pm0.03$ & $20$& $-15.38\pm0.03$ &  $2.22\pm0.08$ &  $1.88\pm0.05$ & $50$\\
  G034-G2(MM2) & -- &  -- & -- & $10$& $42.18\pm0.02$ &  $1.18\pm0.06$ &  $0.46\pm0.02$ & $20$\\
  G034-F1(MM8) & $58.0\pm0.2$ &  $1.4\pm0.3$ &  $0.06\pm0.01$ & $10$& $58.5\pm0.2$ &  $1.2\pm0.2$ &  $0.27\pm0.01$ & $20$\\
  G034-F2(MM7)  &  -- &  -- &  -- & $10$& $56.74\pm0.03$ &  $0.81\pm0.07$ &  $0.15\pm0.01$ & $20$\\
  *G028-C1(MM9) & $80.3\pm0.4$ &  $4\pm1$ &  $0.14\pm0.03$ & $10$& $81.4\pm0.1$ &  $1.6\pm0.2$ &  $0.29\pm0.04$ & $20$\\
  G028-C3(MM11)& -- &  -- &  -- & $10$& $81.32\pm0.09$ &  $1.2\pm0.2$ &  $0.1\pm0.01$ & $20$\\
  I20293-WC & $7.5\pm0.1$ &  $1.3\pm0.2$ &  $0.13\pm0.04$ & $10$& $7.80\pm0.01$ &  $1.26\pm0.02$ &  $0.91\pm0.03$ & $20$\\
  I22134-G & $-18.46\pm0.03$ &  $1.56\pm0.07$ &  $0.39\pm0.01$ & $10$& $-18.13\pm0.02$ &  $1.17\pm0.06$ &  $0.34\pm0.02$ & $20$\\
  I22134-B & $-18.25\pm0.08$ &  $1.3\pm0.2$ &  $0.07\pm0.008$ & $9$& $-18.37\pm0.04$ &  $1.03\pm0.07$ &  $0.12\pm0.08$ & $10$\\
   \midrule
  \multicolumn{9}{c}{HMPO}\\
  \midrule
   I00117-MM1 & $-36.0\pm0.1$ &  $2.1\pm0.2$ &  $0.15\pm0.06$ & $10$& $-35.48\pm0.05$ &  $1.8\pm0.1$ &  $0.26\pm0.02$ & $20$\\
 AFGL5142-MM & $-2.67\pm0.04$ &  $3.3\pm0.1$ &  $1.42\pm0.04$ & $20$& $-1.62\pm0.04$ &  $3.2\pm0.1$ &  $2.55\pm0.08$ & $60$\\
  05358-mm1 & $-16.12\pm0.05$ &  $2.8\pm0.1$ &  $0.80\pm0.03$ & $20$& $-15.62\pm0.03$ &  $2.11\pm0.07$ &  $1.50\pm0.04$ & $40$\\
  18089-1732 & $32.15\pm0.07$ &  $2.8\pm0.2$ &  $0.820\pm0.002$ & $20$& $32.19\pm0.05$ &  $1.9\pm0.1$ &  $0.74\pm0.04$ & $30$\\
  18517+0437 & $44.00\pm0.04$ &  $2.9\pm0.1$ &  $0.89\pm0.03$ & $20$ & $44.37\pm0.04$ &  $2.77\pm0.09$ &  $1.48\pm0.04$ & $40$\\
  G75-core & $0.07\pm0.01$ &  $3.1\pm0.2$ &  $0.79\pm0.09$ & $10$& $0.3\pm0.2$ &  $2.5\pm0.2$ &  $0.28\pm0.01$ & $20$\\
  I20293-MM1 & $6.42\pm0.05$ &  $2.2\pm0.1$ &  $0.44\pm0.02$ & $20$& $6.53\pm0.06$ &  $2.6\pm0.2$ &  $0.76\pm0.04$ & $30$\\
  I21307 & $-46.2\pm0.1$ &  $2.2\pm0.2$ &  $0.13\pm0.01$ & $10$& $-46.29\pm0.08$ &  $1.7\pm0.2$ &  $0.18\pm0.02$ & $20$\\
  I23385 & $-50.1\pm0.01$ &  $2.4\pm0.3$ &  $0.32\pm0.03$ & $10$& $-49.43\pm0.07$ &  $1.2\pm0.2$ &  $0.15\pm0.02$ & $20$\\
    \midrule
  \multicolumn{9}{c}{UC H$_{II}$}\\
  \midrule
   G5.89-0.39 & $8.94\pm0.03$ &  $3.9\pm0.1$ &  $4.4\pm0.2$ & $20$ & $8.06\pm0.01$ &  $1.98\pm0.04$ &  $2.28\pm0.06$ & $30$\\
  I19035-VLA1 & $32.9\pm0.2$ &  $4.7\pm0.4$ &  $0.35\pm0.03$ & $10$& $32.9\pm0.1$ &  $2.6\pm0.3$ &  $0.34\pm0.05$ & $20$\\
  19410+2336 & $22.60\pm0.03$ &  $1.94\pm0.08$ &  $0.66\pm0.02$ & $20$& $22.87\pm0.01$ &  $1.85\pm0.04$ &  $1.31\pm0.02$ & $20$\\
  ON1 & $11.83\pm0.05$ &  $4.1\pm0.1$ &  $1.01\pm0.03$ & $10$& $12.59\pm0.04$ &  $5.11\pm0.09$ &  $1.69\pm0.03$ & $20$\\
  I22134-VLA1 & $-18.34\pm0.03$ &  $1.47\pm0.08$ &  $0.09\pm0.001$ & $10$& $-18.19\pm0.04$ &  $1.72\pm0.08$ &  $0.33\pm0.01$ & $20$\\
  23033+5951 & $-52.93\pm0.06$ &  $2.6\pm0.1$ &  $0.51\pm0.02$ & $20$& $-52.48\pm0.02$ &  $1.11\pm0.06$ &  $0.64\pm0.004$ & $30$\\
  NGC 7538-IRS9 & $-56.82\pm0.08$ &  $4.1\pm0.3$ &  $0.77\pm0.04$ & $20$& $-56.54\pm0.05$ &  $1.9\pm0.1$ &  $0.61\pm0.03$ & $30$\\
  \bottomrule
 \end{tabular}
 \end{center}
 * tentative detection, as explained in Sect.~3.1.
\label{hc15n-dnc-fit}
\normalsize
\end{table}

\begin{table}
\caption{Values obtained from the hyperfine fit of the H$^{13}$CN(1-0) lines. The errors are coming from the fitting procedure and are not taking into account of the calibration error on T$_{MB}$. In the second and in the third colums are listed the centroid velocities and the FWHM, respectively. In the fourth and fifth columns there are the optical depths and the antenna temperatures times the optical depths. In the sixth column are listed the r.m.s of the spectra, and finally in the last column there are the integrated intensities that, in this case, are not obtained from the fitting procedure but from the PRINT AREA command in CLASS.}
  \begin{tabular}{l*{6}{c}}
  \toprule
  Source & $v_{LSR}$ & $\Delta v_{1/2}$ & $\tau$ & $T_{A}\times\tau$ & $\sigma$& $\int T_{MB}\,dv$\\
  &(km/s) & (km/s) &  & (K) & (mK)& (K km/s)\\
  \midrule
  \multicolumn{6}{c}{HMSC}\\
  \midrule
  I00117-MM2 & $-36.37\pm0.05$ &  $1.8\pm0.1$  &  $0.1\pm0.1$ &  $0.142\pm0.008$ & $10$&  $0.42\pm0.02$\\
  AFGL5142-EC & $-3.35\pm0.02$ &  $3.30\pm0.04$  &  $0.10\pm0.06$ &  $0.94\pm0.01$ & $30$&  $5.81\pm0.08$\\
  05358-mm3 & $-16.65\pm0.03$ &  $3.00\pm0.05$  &  $0.10\pm0.04$ &  $0.80\pm0.01$ & $30$&  $4.51\pm0.07$\\
  G034-G2(MM2) & $41.1\pm0.1$ &  $1.7\pm0.2$  &  $0.1\pm0.4$ &  $0.087\pm0.007$ & $10$&  $0.23\pm0.02$\\
  G034-F1(MM8) & $57.62\pm0.08$ &  $1.7\pm0.1$  &  $0.1\pm0.4$ &  $0.16\pm0.01$ & $10$&  $0.50\pm0.02$\\
  G034-F2(MM7) & $55.69\pm0.04$ &  $0.8\pm0.1$  &  $0.25\pm0.02$ &  $0.11\pm0.01$ & $10$&  $0.23\pm0.01$ \\
  G028-C1(MM9) & $79.43\pm0.05$ &  $3.09\pm0.09$  &  $0.1\pm0.2$ &  $0.180\pm0.006$ & $10$&  $1.02\pm0.03$\\
  G028-C3(MM11)& $80.31\pm0.08$ &  $2.2\pm0.3$  &  $0.5\pm0.7$ &  $0.12\pm0.03$ & $10$&  $0.45\pm0.03$\\
  I20293-WC & $6.83\pm0.06$ &  $1.0\pm0.1$   &  $0.60\pm0.03$&  $0.15\pm0.02$ & $10$&  $1.33\pm0.03$\\
  I22134-G & $-19.04\pm0.01$ &  $1.65\pm0.03$   &  $0.10\pm0.03$&  $0.63\pm0.01$ & $10$&  $1.85\pm0.03$ \\
  I22134-B & $-18.95\pm0.02$ &  $1.43\pm0.06$   &  $0.1\pm0.4$&  $0.164\pm0.006$ & $10$&  $0.45\pm0.02$\\
   \midrule
  \multicolumn{6}{c}{HMPO}\\
  \midrule
   I00117-MM1 & $-36.54\pm0.03$ &  $1.96\pm0.09$  &  $0.3\pm0.3$ &  $0.21\pm0.02$ & $10$&  $0.74\pm0.02$\\
 AFGL5142-MM & $-3.43\pm0.01$ &  $3.27\pm0.03$   &  $0.26\pm0.06$ &  $1.27\pm0.03$ & $20$&  $7.52\pm0.05$ \\
  05358-mm1 & $-16.8\pm0.02$ &  $2.99\pm0.04$   &  $0.50\pm0.02$&  $0.91\pm0.01$ & $20$&  $4.68\pm0.06$\\
  18089-1732 & $31.8\pm0.4$ &  $2\pm1$   &  $0.1\pm0.1$ &  $01.48\pm0.02$ & $30$&  $6.51\pm0.05$\\
  18517+0437 & $43.29\pm0.01$ &  $2.97\pm0.02$  &  $0.1\pm0.03$ &  $1.024\pm0.007$ & $20$&  $5.76\pm0.05$\\
  G75-core & $-0.596\pm0.004$ &  $2.5\pm0.1$  &  $0.1\pm0.02$ &  $0.48\pm0.02$ & $10$&  $3.27\pm0.02$ \\
  I20293-MM1 & $5.79\pm0.02$ &  $2.38\pm0.03$  &  $0.100\pm0.002$ &  $0.80\pm0.01$ & $20$&  $3.64\pm0.04$ \\
  I21307 & $-47.00\pm0.04$ &  $2.3\pm0.1$  &  $0.2\pm0.3$ &  $0.18\pm0.02$ & $10$&  $0.75\pm0.03$\\
  I23385 & $-50.9\pm0.3$ &  $1\pm1$  &  $0.6\pm0.1$ &  $0.26\pm0.01$ & $10$&  $1.16\pm0.02$\\
    \midrule
  \multicolumn{6}{c}{UC H$_{II}$}\\
  \midrule
   G5.89-0.39 & $8.24\pm0.01$ &  $3.83\pm0.01$  &  $0.12\pm0.01$ &  $3.75\pm0.2$ & $30$&  $20.97\pm0.05$\\
  I19035-VLA1 & $32.17\pm0.05$ &  $4.4\pm0.1$  &  $0.4\pm0.2$ &  $0.30\pm0.02$ & $10$&  $2.26\pm0.04$\\
  19410+2336 & $22.01\pm0.002$ &  $2.186\pm0.002$  &  $0.10\pm0.02$ &  $1.090\pm0.005$ & $20$&  $4.64\pm0.04$ \\
  ON1 & $11.12\pm0.02$ &  $3.71\pm0.04$  &  $0.73\pm0.08$ &  $0.93\pm0.03$ & $10$&  $5.55\pm0.04$\\
  I22134-VLA1 & $-18.98\pm0.01$ &  $1.76\pm0.03$  &  $0.10\pm0.03$ &  $0.513\pm0.007$ & $10$&  $1.67\pm0.03$\\
  23033+5951 & $-53.63\pm0.02$ &  $2.73\pm0.02$  &  $0.10\pm0.02$ &  $0.621\pm0.003$ & $20$&  $3.17\pm0.04$\\
  NGC 7538-IRS9 & $-57.64\pm0.02$ &  $3.60\pm0.04$  &  $0.10\pm0.03$ &  $0.65\pm0.01$ & $20$&  $4.43\pm0.04$\\
  \bottomrule
 \end{tabular}
\centering
\label{h13cn-fit}
\end{table}

\clearpage

\section{Spectra}
In this appendix, all spectra of HN$^{13}$C(1-0), H$^{13}$CN(1-0), H$^{15}$NC(1-0), HC$^{15}$NC(1-0) and DNC(2-1) transitions for all the sources are shown.
\clearpage



\clearpage
\section{Comparison with other molecules}
In this appendix, we compare the $^{14}$N/$^{15}$N ratios obtained in this paper and in Fontani et al.~(\citeyear{fontani15a}) for all the common sources. The $^{12}$C/$^{13}$C used and the Galactocentric distances are also listed.
\clearpage
\begin{table*}
\begin{center}
 \caption{{ In the second and in the third columns are listed the $^{14}$N/$^{15}$N ratios for HNC and HCN found in this work. In the fourth, fifth and sixth columns are listed the $^{14}$N/$^{15}$N ratios for N$_{2}$H$^{+}$ and CN found by Fontani et al.~(\citeyear{fontani15a}). In the seventh column are listed the values of $^{12}$C/$^{13}$C found from the gradient with the Galactocentric distances given by Milam et al.~(\citeyear{milam05}). In the last column are listed the Galactocentric distances of the sources obtained from the source distance from the Sun (taken from Fontani et al.~\citeyear{fontani11}) and the corresponding galactic longitude.}}
\label{comparisondifferentmolecules}
\begin{tabular}{llllllll}
 \hline \hline
 Source &$\frac{HNC}{H^{15}NC}$ &$\frac{HCN}{HC^{15}N}$ &$\frac{N_{2}H^{+}}{^{15}NNH^{+}}$ &$\frac{N_{2}H^{+}}{N^{15}NH^{+}}$ & $\frac{CN}{C^{15}N}$ & $\frac{^{12}C}{^{13}C}$ & D$_{GC}$ \\
          & && &  &   &  & (kpc) \\
 \cline{1-8}
 \multicolumn{8}{c}{HMSC} \\
 \cline{1-8}
 I00117-MM2 &  $460\pm 80$ & $282\pm54$ &$670\pm98$ &$\geq561$ & -- & $69$& $9.5$\\
 AFGL5142-EC & $398\pm39$&$398\pm18$ &$1100\pm360$ &$\geq1034$ & $330\pm67$& $74$& $10.3$\\
  05358-mm3 &  $388\pm28$&$433\pm20$ &$210\pm12$&$180\pm13$ & $400\pm90$& $74$& $10.3$\\
  G034-G2(MM2) & $365\pm71$&$\geq147$& $\geq856$ & $\geq873$& -- &$50$ &$6.3$ \\
  G034-F1(MM8)  &  $495\pm72$&$369\pm80$ &$\geq566$& $\geq672$& -- &$47$ & $5.8$\\
  G034-F2(MM7)   & $783\pm263$ &$\geq 168$& $\geq195$ & $\geq232$& -- & $47$&$5.8$\\
  G028-C1(MM9) & $260\pm30$ &$287\pm55$& $\geq1217$& $\geq1445$& -- & $40$& $4.7$ \\
  G028-C3(MM11)&  $338\pm53$ &$\geq250$ & -- & -- & -- &$40$ & $4.7$ \\
  I20293-WC  & $439\pm90$  &$620\pm180$ & $65\pm69$&$550\pm98$ & -- & $62$& $8.3$\\
  I22134-G  &$369\pm35$  &$322\pm14$& $\geq232$& $\geq197$& $330\pm150$&$69$ & $9.5$\\
  I22134-B   & $414\pm85$  &$406\pm50$ &$\geq316$ & $\geq322$& -- & $69$&$9.5$\\
  \cline{1-8}      
\multicolumn{8}{c}{HMPO} \\
\cline{1-8}
   I00117-MM1  &  $388\pm75$&$345\pm38$ &$220\pm60$ & $\geq233$& $\geq250$& $69$&$9.5$\\
 AFGL5142-MM  &$433\pm27$ &$387\pm10$& $740\pm97$& $1300\pm210$& $190\pm23$&$74$ & $10.3$\\
  05358-mm1   & $466\pm29$ &$432\pm18$&$190\pm20$ &$180\pm23$ & $240\pm40$& $74$& $10.3$\\
  18089-1732   &$385\pm204$ &$338\pm9$ &$1000\pm400$ & $800\pm100$& $450\pm100$& $43$& $5$\\
  18517+0437 & $349\pm22$ &$326\pm9$ &$390\pm53$ & $260\pm20 $& $310\pm80$&$51$ & $6.4$\\
  G75-core & $554\pm84$ &$258\pm3$ & $\geq209$&$\geq240$ & $240\pm50$&$63$ & $8.4$\\
  I20293-MM1   & $481\pm33$& $504\pm25$ &$790\pm67$ &$370\pm42$ & -- &$62$ & $8.3$\\
  I21307  & $\geq317$&$389\pm46$ & $\geq301$&$\geq346$ & $\geq175$& $68$& $9.3$\\
  I23385& $639\pm189$ &$290\pm26$ &$\geq81$ & $\geq66$& -- & $81$& $11.4$\\
    \cline{1-8}													      	  
\multicolumn{8}{c}{UC HII} \\
\cline{1-8}
G5.89-0.39  & $432\pm10$&$255\pm12$ & $320\pm25$&$450\pm51$ & $350\pm160$& $55$& $7.2$\\
  I19035-VLA1  & $570\pm107$&$340\pm27$ & $\geq572$&$\geq754$ & $\geq270$& $54$& $7$\\
  19410+2336 & $431\pm24$ &$405\pm12$& $450\pm96$ &$500\pm100$ &$430\pm250$ & $58$& $7.7$\\
  ON1   & $467\pm19$&$327\pm8$ &$460\pm22$ &$350\pm33$ & $350\pm220$& $60$& $8$ \\
  I22134-VLA1  &  $364\pm43$ &$390\pm18$ &$\geq80$&$\geq108$ & $250\pm84$& $69$& $9.5$ \\
  23033+5951  & $397\pm35$  &$458\pm21$& $900\pm400$&$700\pm250$ & $\geq370$& $74$& $10.2$\\
  NGC 7538-IRS9 & $331\pm37$  &$410\pm20$ & $\geq255$& $\geq294$& $230\pm90$& $72$& $9.9$\\
\hline
\end{tabular}
\label{alldata}
\end{center}
\end{table*}
\normalsize
\end{appendix}
\end{document}